\documentclass[sigplan,screen]{acmart}

\usepackage{hyperref}

\usepackage{pifont}
\newcommand{\xmark}{\ding{55}}%
\usepackage{amsmath,amsfonts}
\usepackage{algorithmic}
\usepackage{tikz}
\usepackage{graphicx}
\usepackage{textcomp}
\usepackage{xcolor}
\usepackage{booktabs}
\usepackage{siunitx}
\usepackage{algorithm2e}
\usepackage{multirow}
\usepackage{fancyhdr}
\usepackage{float}
\usepackage{setspace}
\usepackage{listings}
\usepackage{balance}
\usepackage{wrapfig}
\usepackage{caption}
\usepackage{boldline} %
\usepackage{colortbl} %
\usepackage{floatflt}
\usepackage{tabularray}
\usepackage{rotating}
\usepackage{eso-pic}
\usepackage{titlesec}
\usepackage{noindentafter}
\usepackage{datetime}
\usepackage[]{hyphenat}
\usepackage{tikzducks}
\usepackage{makecell}
\usepackage[absolute,overlay]{textpos}
\usepackage{everypage} %
\usepackage{balance}
\usepackage{xspace}
\usepackage{color}
\usepackage{listings}
\usepackage{xcolor}
\usepackage[most]{tcolorbox}
\usepackage{framed}
\usepackage{mdframed}

\definecolor{shadecolor}{gray}{0.9} %
\definecolor{uglyyellow}{rgb}{0.99, 0.93, 0.0}
\definecolor{akturquoise}{rgb}{0.04,0.59,0.46}
\definecolor{nbs}{rgb}{0.35, 0.31, 0.81}

\newcommand{\nb}[1]{\textcolor{black}{#1}}
\newcommand{\ak}[1]{\textcolor{black}{#1}}

\newcommand{\bkkc}[1]{\todo[size=\scriptsize, linecolor=black, bordercolor=black, backgroundcolor=white]{}}
\definecolor{darkred}{rgb}{0.75, 0.1, 0.1}

\newcommand{\good}[1]{\textcolor{darkspringgreen}{#1}}
\newcommand{\bad}[1]{\textcolor{red}{#1}}

\usepackage{tikz}

\newcommand{\squishlist}{
 \begin{list}{$\circ$}
  { \setlength{\itemsep}{0pt}
     \setlength{\parsep}{0pt}
     \setlength{\topsep}{3pt}
     \setlength{\partopsep}{0pt}
     \setlength{\leftmargin}{1em}
     \setlength{\labelwidth}{1em}
     \setlength{\labelsep}{0.5em} } }

\newcommand{\squishend}{
  \end{list}  }

\newcommand*\circled[1]{\tikz[baseline=(char.base)]{\node[shape=circle,fill,inner sep=0.5pt] (char) {\textcolor{white}{#1}};}}
\newcommand*\circledwhite[1]{\tikz[baseline=(char.base)]{\node[shape=circle, inner sep=0.5pt, draw=black, fill=white, text=black] (char) {#1};}}

\usepackage[bottom]{footmisc}
\usepackage{dblfloatfix}
\usepackage{array}
 \usepackage{booktabs}

\newcommand\head[1]{{\noindent\textbf{#1}.}}
\definecolor{ufogreen}{rgb}{0.1, 0.6, 0.4}
\definecolor{ufogreen}{rgb}{0.1, 0.6, 0.4}

\newcommand{\power}{0.72}

\usepackage{listings}
\usepackage{xcolor}
\usepackage[T1]{fontenc} %
\usepackage{zi4} %
\usepackage{caption} %
\usepackage{float} %

\lstdefinestyle{custompseudocode}{
  belowcaptionskip=1\baselineskip,
  breaklines=true,
  xleftmargin=\parindent,
  language=Python,
  showstringspaces=false,
  basicstyle=\small\ttfamily,
  keywordstyle=\bfseries\color{green!40!black},
  commentstyle=\itshape\color{purple},
  stringstyle=\color{orange},
  numbers=left,
  numberstyle=\scriptsize\color{black},
  numbersep=8pt,
  morekeywords={function}, %
  keywordstyle=[2]\bfseries, %
  escapeinside={*@}{@*}, %
  xleftmargin=2em, xrightmargin=0em, 
}
\DeclareCaptionFormat{listing}{%
  \rule{\linewidth}{0.5pt}
  \vspace{-1mm}
  \textbf{#1#2: }#3
  \vspace{-1mm}
  \rule{\linewidth}{0.5pt}
  \vspace{-7mm}
}

 \definecolor{chocolate}{rgb}{0.92, 0.41, 0.12}
\definecolor{burgundy}{rgb}{0.5, 0.0, 0.13}
\definecolor{darkmagenta}{rgb}{0.55, 0.0, 0.55}
\definecolor{darkblue}{rgb}{0.0, 0.5, 1.0}

\definecolor{darkspringgreen}{rgb}{0.09, 0.45, 0.27}
\definecolor{denim}{rgb}{0.08, 0.38, 0.74}
\definecolor{darkolivegreen}{rgb}{0.33, 0.42, 0.18}
\definecolor{tangerine}{rgb}{0.99, 0.52, 0.3}
\definecolor{mahogany}{rgb}{0., 0.25, 0.0}
\definecolor{brown}{rgb}{0.47, 0.33, 0.28}
\definecolor{pale}{rgb}{0.97, 0.96, 0.88}
\definecolor{darkorange}{rgb}{0.49, 0.30, 0.08}

\newcommand\sepherd[1]{\textcolor{black}{#1}}
\newcommand\konrevia[1]{\textcolor{black}{#1}}

\newcommand\konrevic[1]{\textcolor{black}{#1}}
\newcommand\konrevid[1]{\textcolor{black}{#1}}
\newcommand\konrevie[1]{\textcolor{black}{#1}}
\newcommand\konrevif[1]{\textcolor{black}{#1}}

\newcommand\akdel[1]{\textcolor{black}{#1}}

\newcommand\konrevb[1]{\textcolor{black}{#1}}
\newcommand\konrevc[1]{\textcolor{black}{#1}}

\newcommand\konreve[1]{\textcolor{black}{#1}}

\newcommand\konrevatodo[1]{}
\newcommand\konrevbtodo[1]{}
\newcommand\konrevctodo[1]{}
\newcommand\konrevdtodo[1]{}
\newcommand\konrevetodo[1]{}
\newcommand\konrevftodo[1]{}

\hypersetup{
    bookmarks=true,
    breaklinks=true,
    colorlinks=true,
    linkcolor=blue,
    citecolor=blue,
    urlcolor=black,
    pdfpagelabels=false,
}

\newcommand{\obslbl}[0]{Obsv.}

\newcommand\observation[1]{%
   \noindent
   \colorbox{gray!20}{\textbf{\obslbl{}}} \emph{#1}}

\newcommand\VMcharacterization{~\cite{vm2,karakostas_characterIISWC,
isca2010-barr-trancache,5-levelpaging,contiguitas2023,radiantISMM21,bhattacharjeePACT2009,devirtualizingASPLOS2018,hash_dont_cache,virtualizationimplication,vm29,
vmhpcIISWC2024nick}}

\newcommand\VMlargepages{~\cite{park2020perforated,guvenilir2020tailored,ingensOSDI2016,talluriISCA1992,panwar2018making,panwar2019hawkeye,tridentMICRO2021,pham2015,mosaic2017MICRO,promotionHPCA2001,shadowpageISCA1998,duHPCA2015,vm42,vm43,partialMICRO2020,ganapathy98}}

\newcommand\VMcontiguity{~\cite{translationranger2019,karakostas2015,chloe2020,hybridtlbISCA2017,flexpointerTACO2023,contiguitas2023,vm6,vm2,dmtASPLOS2024}}

\newcommand\VMpwcs{~\cite{isca2010-barr-trancache,vm10,esteve14}}

\newcommand\VMpagetable{~\cite{haria2018, flataAsplos2022,elastic-cuckoo-asplos20,mehtJovanHPCA2023,hash_dont_cache,impicaICCD2016,mitosis-asplos20, compendiaISMM2021, Alam2017DoItYourselfVM,distributedptMICRO24osang}}

\newcommand\VMvirtualized{~\cite{vm25,vm35,pham2015,pham2015tr,vm11,virtcoherenceISCA2017,babelfish,margaritov2021ptemagnet,virtcoherenceISCA2017,panwar2021fast}}

\newcommand\VMintermediate{~\cite{enigma,midgard,vbi,powerpc2003}}

\newcommand\VMtlbprefetching{~\cite{vavouliotis2021,morriganMICRO2021,margaritov2019prefetched,kandiraju2002going,saulsbury2000recency,Bala1994SoftwarePA}}

\newcommand\VMtlbreplacementpolicy{~\cite{deadTLBHPCA2021,chirpMICRO2020}}

\newcommand\VMrestrictive{~\cite{kanellopoulos2023utopia,nearmemoryPact17,mosaicpagesASPLOS2023}}

\newcommand\VMsoftwareTLB{~\cite{pomtlbISCA2017,csaltMICRO2017,softwareTLBNAS2013,softwareTLBISCA2013,uhlig94,bruceMMU1998,softcontrolcachesISCA1986,Nagle1993DesignTF,Bala1994SoftwarePA}}

\newcommand\VMtlbincache{~\cite{kanellopoulosMICRO2023victima,gputlbreachMICRO2021,ducati}}

\newcommand\VMmethodology{~\cite{hybridtlbISCA2017, chloe2020, directsegments, Karakostas2015Redundant, hash_dont_cache, margaritov2019prefetched, papadopoulou2015,compendiaISMM2021, multiplesizesASPLOS2017, pham2015, spectlbISCA2011}}

\newcommand\VMold{~\cite{ieemicro2018-Bhattacharjee-tempo,hand1999,old_vm1,old_vm2,old_vm3,old_vm4,old_vm5,old_vm6,old_vm7,denning1970,ahearn1973,goldberg1974survey,bruceMMU1998,smith,wood1986,chen1992simulation,koldinger1992,lindstrom1995,jacob1998,avm,translationmanagementISCA1993,interactionASPLOS1991,multics}}

\newcommand\SimFPGA{~\cite{ramp2006,fastMICRO2007,hasim,protoflex,ramp-gold,firesim}}

\newcommand\GPUUVM{~\cite{wangoasis,wanggrit}}

\newcommand\HybridNVM{~\cite{li2017utility,zhao2014firm,salkhordeh2019analytical,meza2012enabling, janus, daxvm}}

\newcommand\MemoryCXL{~\cite{rethinkingISMM22,lowOverheadCXL,demystifyingCXLType2,sun2023demystifying,masouros2023adrias,guo2022clio, Gao2016, Shan2018, Korolija2021, Wang2020, Zuo2021, Maruf2020, Lim2009, Zhang2020, Yan2019, Angel2020, Lim2012, Peng2020, Bindschaedler2020, Katrinis2016, Aguilera2017, Aguilera2018, Rao2016, Calciu2021, Adya2019, Lagar2019, Pinto2020, Gu2017, Buragohain2017, Zervas2018}}

\newcommand\Workloads{~\cite{gcn2016,gnnSurvey,memcached1,redis2,graph500,graphaligner2020,cali2022segram,luszczek_hpcc2006,Tramm2014,Lifeng2015,recommender,recommender1}}

\newcommand\VMworkloads{~\cite{gcn2016,gnnSurvey,memcached1,redis2,graph500,graphaligner2020,cali2022segram,luszczek_hpcc2006,Tramm2014,Lifeng2015,recommender,recommender1,vllmSOSP2023,jiang2023mistral7b,murty2024bagelbootstrappingagentsguiding,touvron2023llamaopenefficientfoundation,Ustiugov_2021,igniteMICRO23,mementoMICRO2023,catalyzerASPLOS2020,implicationsfaasMICRO2019,deathstarbench}}

\newcommand\Allsimulators{~\cite{simplescalar,multi2sim,scarab,luo2023ramulator2,sanchez2013zsim,gem5,champsim,sniper,ptlsim,qflexv2.0,gem5,Hardavellas2004SimFlexAF, graphite, simics, dracksim, marss, cmpsim}}

\newcommand\Emulationbased{~\cite{simplescalar,multi2sim,scarab,luo2023ramulator2,sanchez2013zsim,gem5,champsim,sniper,dracksim, cmpsim}}

\newcommand\Fullsystem{~\cite{Hardavellas2004SimFlexAF,ptlsim,qflexv2.0,gem5,graphite,simics,marss}}

\newfloat{listing}{tbhp}{lst}[section] %
\floatname{listing}{Listing}
\copyrightyear{2025}
\acmYear{2025}
\setcopyright{acmlicensed}
\acmConference[ASPLOS '25]{Proceedings of the 30th ACM International Conference on Architectural Support for Programming Languages and Operating Systems, Volume 2}{March 30-April 3, 2025}{Rotterdam, Netherlands}
\acmBooktitle{Proceedings of the 30th ACM International Conference on Architectural Support for Programming Languages and Operating Systems, Volume 2 (ASPLOS '25), March 30-April 3, 2025, Rotterdam, Netherlands}
\acmDOI{10.1145/3676641.3716027}
\acmISBN{979-8-4007-1079-7/2025/03}

\begin{document}

\author{
  Konstantinos Kanellopoulos\textsuperscript{1}\quad\quad
  Konstantinos Sgouras\textsuperscript{1}\quad\quad
  F. Nisa Bostanci\textsuperscript{1}\quad\quad
  \newline Andreas Kosmas Kakolyris \textsuperscript{1}\quad\quad
  Berkin Kerim Konar \textsuperscript{1}\quad\quad
  Rahul Bera\textsuperscript{1}\quad\quad 
  \newline Mohammad Sadrosadati\textsuperscript{1}\quad\quad 
  Rakesh Kumar \textsuperscript{2}\quad\quad
  Nandita Vijaykumar \textsuperscript{3}\quad\quad
  Onur Mutlu\textsuperscript{1}
  \newline
}

\affiliation{\textsuperscript{1}ETH Zürich\quad \textsuperscript{2}Norwegian University of Science and Technology \quad  \textsuperscript{3}University of Toronto}
\email{}
\thanks{}
\renewcommand{\shortauthors}{Kanellopoulos et al.}
\title{ Virtuoso: Enabling Fast and Accurate \\ Virtual Memory Research via an Imitation-based \\ Operating System Simulation Methodology}
\renewcommand{\shorttitle}{Virtuoso: Enabling Fast and Accurate Virtual Memory Research via \\ an Imitation-based  Operating System Simulation Methodology}

\renewcommand{\authors}{Konstantinos Kanellopoulos,
Konstantinos Sgouras,
F. Nisa Bostanci,
Andreas Kosmas Kakolyris,
Berkin Kerim Konar,
Rahul Bera,
Mohammad Sadrosadati,
Rakesh Kumar,
Nandita Vijaykumar,
Onur Mutlu}

\begin{abstract}
    The unprecedented growth in data demand from emerging applications has turned virtual memory (VM) into a major performance bottleneck. VM's overheads are expected to persist as memory requirements continue to increase. 
    Researchers explore new hardware/OS co-designs to optimize VM across diverse applications and systems. 
    To evaluate such designs, researchers rely on various simulation methodologies to model VM components.
    Unfortunately, current simulation tools (i) either lack the desired accuracy in modeling VM's software components or (ii)  are too slow and complex to prototype and evaluate schemes that span across the hardware/software boundary.
    
    We introduce Virtuoso, a new simulation framework that enables quick and accurate prototyping and evaluation of the \textit{software and hardware components} of the VM subsystem. 
    The key idea of Virtuoso is to employ a {\emph{lightweight userspace OS kernel}}, called MimicOS, that (i) accelerates simulation time by imitating \emph{only} the desired kernel functionalities, (ii)  facilitates the development of new OS routines that imitate real ones, using an accessible high-level programming interface, (iii) enables accurate and flexible evaluation of the application- and system-level implications of VM after integrating {Virtuoso} to a desired architectural simulator. 

    {In this work, we integrate Virtuoso into five diverse architectural simulators, each specializing in different aspects of system design, and heavily enrich it with multiple state-of-the-art VM schemes. This way, we establish a common ground for researchers to evaluate current VM designs and to develop and test new ones. We demonstrate Virtuoso's flexibility and versatility by evaluating five diverse use cases, yielding new insights into state-of-the-art VM techniques.} Our validation shows that Virtuoso ported on top of Sniper, a state-of-the-art microarchitectural simulator, models (i) the memory management unit  of a real high-end server-grade CPU with 82\% accuracy, and (ii) the page fault latency of a real Linux kernel with up to 79\% accuracy. Consequently, Virtuoso models the IPC {performance} of a real high-end server-grade CPU with {21\%} higher accuracy than the {baseline} version of Sniper. Virtuoso's accuracy benefits incur an average simulation time overhead of only 20\%, {on top of four baseline architectural simulators.}
    The source code of Virtuoso is freely available at \textcolor{blue}{\url{https://github.com/CMU-SAFARI/Virtuoso}}.

    \end{abstract}

\maketitle

\vspace{-3mm}
\section{Introduction}

\konrevia{Virtual memory (VM)\VMold~} is a cornerstone of modern computing systems, enabling \konrevia{application-}transparent \konrevia{physical} memory management, isolation and data sharing. 
\konrevia{Contemporary applications (e.g., \VMworkloads) exhibit different characteristics that stress the VM subsystem. 
We classify these workloads into two broad categories: (i) long-running workloads (i.e., execution time larger than 100s of seconds)~\cite{graph500,recommender,recommender1,graphaligner2020,cali2022segram,luszczek_hpcc2006,gcn2016,Lifeng2015} with large data footprints and irregular memory access patterns, that exhibit high address translation overheads,
and (ii) short-running workloads (i.e., execution time often lower than 1 second)~\cite{vllmSOSP2023,jiang2023mistral7b,murty2024bagelbootstrappingagentsguiding,touvron2023llamaopenefficientfoundation,Ustiugov_2021,igniteMICRO23,mementoMICRO2023,catalyzerASPLOS2020,implicationsfaasMICRO2019,deathstarbench} whose execution time does not amortize the overheads of system software operations (e.g., physical memory allocation).}
Multiple prior works and industrial studies\VMcharacterization~have shown that
address translation in long-running workloads and memory allocation in short-running workloads respectively account  
for up to 40\% and 95\% of the total execution time. 
As memory requirements continue to \konrevia{increase} and systems transition to larger physical address spaces~\cite{intelicelake} 
(e.g., \konrevia{via} hybrid memory systems with high-capacity non-volatile memories\HybridNVM, \konrevia{memory disaggregation\MemoryCXL})\konrevia{,} the overheads associated with VM operations are expected to increase.

To tackle these overheads, many research works take a hardware/OS co-design approach and revisit core aspects of VM such as page table structure\VMpagetable, %
virtual-to-physical mapping~\cite{rmmisca15, kanellopoulos2023utopia, vbi, mosaicpagesASPLOS2023, directsegments,nearmemoryPact17,enigma,midgard},  
physical memory \konrevia{allocation} policy (e.g. transparent huge page mechanisms~\cite{ingens,translationranger2019,corbet2011,corbet2017,tridentMICRO2021,psomadakis2024elastic,tridentMICRO2021}) and Translation Lookaside Buffer (TLB) design\VMsoftwareTLB{}.
\konrevia{Evaluating} such VM designs is not straightforward. The evaluation challenge 
primarily arises from the need to model the interplay between \konrevia{both} the OS and HW components involved in VM.
For example, in modern systems, the OS manages the allocation of large pages, which directly affects the effectiveness of the TLB~\cite{guvenilir2020tailored,papadopoulou2015,hybridtlbISCA2017}, memory footprint of the page table (PT),
 PT walk latency~\cite{elastic-cuckoo-asplos20, mehtJovanHPCA2023, distributedptMICRO24osang} and latency of page faults~\cite{ingens,translationranger2019,chloe2020,rmmisca15,minorfaultTACO2022,mementoMICRO2023}. Given this complex interplay, evaluating the strengths and weaknesses of existing and future VM designs becomes a \konrevia{challenging} task without a comprehensive and robust simulation infrastructure.

Unfortunately, modern simulators are either (i) designed for different purposes (e.g., \konrevia{mainly} focus on core microarchitecture~\cite{scarab,champsim,sniper,simplescalar,multi2sim}) \konrevia{and thus lack} the ability and flexibility to accurately model the impact of the OS components involved in the VM subsystem (e.g. Sniper~\cite{sniper}) or (ii) are \konrevia{relatively} slow and hard-to-develop (e.g., gem5 full-system execution mode~\cite{gem5}), \konrevia{which hinders} rapid design space exploration.
This \konrevia{dichotomy of simulators} creates a significant gap in the field, compelling researchers to invest considerable time and effort in developing new \konrevia{custom} tools or methodologies for each VM proposal\VMmethodology.

\textbf{Existing Simulation Methodologies.} Many simulators \konrevic{(e.g., \Emulationbased)} are primarily designed to \konrevic{focus on and} model microarchitectural CPU features. \konrevic{These simulators} emulate basic OS functionalities and \konrevia{use} simplified 
methods to estimate the implications of OS routines on performance. We classify these simulators as \konrevic{\emph{emulation-based}}. Emulation-based simulators often employ
first-order approximations (e.g., fixed latencies) for OS routines and VM operations. \konrevia{As} \ak{we show} in \S\ref{sec:motivation}, %
\ak{fixed} latencies can lead to inaccurate \konrevia{estimation} of VM overheads\konrevia{,} which display high variability across diverse workloads and system states. 
Hence, these simulators \ak{are not suitable \konrevic{for} (and are not primarily designed to be \konrevic{used for})} the evaluation of new VM designs that rely on hardware/OS \ak{co-design}. 
On the other hand, \konrevia{full-system} simulators like gem5~\cite{gem5} and QFlex~\cite{qflexv2.0} allow for detailed 
simulation of the entire OS, supporting realistic memory management for evaluating 
new VM architectures. However, such simulators suffer from significant drawbacks, including 
(i) low simulation \konrevia{speed}, (ii) high memory consumption \konrevia{overhead}, and (iii) 
substantial development \konrevia{effort}. These drawbacks impede \emph{rapid prototyping} of new VM schemes that rely on HW/OS co-design.

As we show in Table~\ref{tab:simulators_comp}, \textbf{our goal} in this work, is to design a simulation framework that (i) \ak{maintains} the speed of emulation-based simulators \ak{while reaching} the accuracy of full-system simulators
and (ii) enables researchers to easily develop and evaluate new VM schemes. 
\konrevic{To this end}, we present \emph{Virtuoso}, a new simulation framework that enables \ak{\konrevia{fast} and accurate prototyping and evaluation of the} \textit{software and hardware components} of the VM subsystem. 
The key idea of Virtuoso is to employ a lightweight userspace kernel, written in a high level language (e.g., C++~\cite{cplusplus}), that enables researchers to (i) isolate the functionality of only the desired kernel code (e.g., \konrevia{Transparent Huge Pages}~\cite{corbet2011,corbet2017}) to speed up simulation time,
(ii) easily develop new OS routines (e.g., \konrevia{a} modified physical \konrevia{memory allocator~\cite{translationranger2019,ingens,chloe2020,psomadakis2024elastic}}) without being kernel experts, and (iii) accurately evaluate \ak{the} application- and system-level implications of \ak{the} OS by integrating \konrevia{Virtuoso} into \konrevia{an} architectural simulator.

\begin{table}[h]
    \centering
    \scriptsize
    \renewcommand{\arraystretch}{1.2} %
    \arrayrulecolor[HTML]{333333} %

    \begin{tabular}{|c|c|c|c|c|}
    \hline
    \rowcolor[HTML]{EFEFEF} %
    \textbf{Simulator-Type} & \textbf{OS} & \textbf{Speed} & \textbf{Accuracy} & \textbf{Development Effort} \\ \hline
    \texttt{Emulation-based} & \bad{N/A} & \good{Fast} & \bad{Low} & \good{Low} \\ \hline
    \texttt{Full-system} & \good{Realistic} & \bad{Slow} & \good{Very High} & \bad{High} \\ \hline
    \hline
    \textbf{\texttt{Our methodology}} & \textbf{\good{Imitation}} & \textbf{\good{Fast}} & \textbf{\good{High}} & \textbf{\good{Low}} \\ \hline
    \end{tabular}
    \vspace{1em}
    \caption{Comparison of existing VM simulation methodologies \konrevia{versus} our proposed methodology for VM research.}
    \label{tab:simulators_comp}
    \vspace{-3mm}
\end{table}

\textbf{Our proposed methodology} involves dynamically instrumenting a userspace kernel that operates as a standalone program and communicates with an architectural simulator via two distinct channels: a functional channel and an instruction stream channel.
The functional channel \konrevia{uses} shared memory primitives and specialized ISA instructions to enable message exchanges between the kernel and the simulator for functional events (e.g., interrupts). For instance, when the simulator triggers a page fault, it communicates this event to the kernel. The kernel then handles the fault and reports the outcome back to the simulator using the shared memory region. Using the instruction stream channel, the kernel injects dynamically instrumented instruction streams (e.g., page fault handler instructions) into the simulator, enabling \konrevia{the simulator} to accurately \konrevia{model} the overheads introduced by OS routines (e.g additional latency, memory interference).

Using this methodology we build MimicOS, a lightweight userspace kernel written in C++~\cite{cplusplus} that imitates, but is not limited to, the basic memory management functionality of the Linux kernel~\cite{linuxkernel}. MimicOS is portable and \konrevia{can be easily attached to the memory model of an architectural simulator} (see \S\ref{sec:integration-other}).
In this work, we integrate MimicOS with \konrevia{five} architectural simulators, Sniper~\cite{sniper}, ChampSim~\cite{champsim}, Ramulator2~\cite{ramulator,luo2023ramulator2}, gem5-SE~\cite{gem5} and an SSD simulator, MQSim~\cite{tavakkol2018mqsim}. 
Using MimicOS and Sniper as a baseline, we build \konrevia{VirTool}, a comprehensive toolset that contains both the HW \konrevia{and} SW components that are required to evaluate many
state-of-the-art VM schemes. By doing so, we aim to (i) unlock a wide range of new case studies ranging from low-level microarchitectural VM schemes \konrevia{to} system software-level ones, and (ii) establish a common ground for researchers to evaluate current VM designs and to develop and test new ones.
Table~\ref{tab:simulators} provides a comprehensive overview of \konrevia{existing} techniques that are included in VirTool.

\textbf{Validation \& Comparison.} We validate the accuracy of MimicOS+Sniper against a real high-end server-grade processor (see \S\ref{sec:validation}) and \konrevia{demonstrate} four key results.
First, MimicOS+Sniper estimates the average L2 TLB \konrevia{misses per kilo instructions} 
and PT walk latency, respectively, with 82\% and 85\% accuracy compared to the real system. 
Second, MimicOS+Sniper estimates the page fault latency with 66\% (up to 79\%) accuracy compared to the page fault latency measured by the Linux kernel running on a real machine. 
Third, MimicOS+Sniper improves \konrevia{instructions per cycle (IPC)} \konrevic{performance estimation} accuracy by \konrevid{21\%} \konrevia{(from 66\% to 80\%)}
while incurring 35\% simulation time overhead compared to baseline Sniper. 
Fourth, MimicOS incurs only 20\% simulation time overhead, \konrevia{averaged} across four simulators, while enabling the full-system execution mode in gem5 leads to 77\% simulation time overhead compared to gem5's system call emulation mode.

\textbf{\konrevia{Versatility} \& Use Cases.} To illustrate the versatility of Virtuoso, we \konrevia{conduct} five case studies that are time\hyp{}consuming and difficult to assess accurately and rapidly using existing simulation tools.
First, we analyze the performance of four different page table designs~\cite{elastic-cuckoo-asplos20,hash_dont_cache} and draw key insights about their impact on page table walk latency, minor page fault latency and main memory interference (see \S\ref{sec:usecase1}).  
Second, we evaluate the overheads associated with different physical memory allocation policies across large language model inference workloads (see \S\ref{sec:usecase2}).
Third, we draw key insights about the architectural trade-offs of restricting the virtual-to-physical address mapping across physical memory~\cite{kanellopoulos2023utopia} (see \S\ref{sec:usecase3}). 
Fourth, we evaluate the benefits of contiguity-aware address translation~\cite{karakostas2015} across different memory fragmentation levels (see \S\ref{sec:usecase4}).
Fifth, we analyze the implications of employing an \konrevia{intermediate address space} scheme~\cite{midgard} across workloads with different memory allocation patterns (see \S\ref{sec:usecase5}).

In this work, we make the following contributions:

\begin{itemize}         
    \item We propose Virtuoso, a new simulation framework that employs a \konrevic{new} imitation-based OS simulation methodology. \konrevic{Virtuoso} enables \konrevia{fast and accurate prototyping and evaluation of} the hardware and  software components of the \konrevic{virtual memory (VM)} subsystem.
    \item We \konrevia{integrate our \konrevic{new} methodology} with five diverse architectural simulators and \konrevic{implement} a comprehensive set of state-of-the-art VM techniques to provide a common ground for researchers to evaluate current and new VM designs.
    \item \konrevia{We validate Virtuoso against a real \konrevic{CPU} system and demonstrate that it improves the accuracy of a \konrevic{state-of-the-art} emulation-based  simulator with \konrevic{only a} modest increase in simulation time.} \konrevic{We demonstrate that Virtuoso can bridge the gap between emulation-based and full-system simulators enabling accurate exploration of VM designs at a fast and flexible way.} 
    \item \konrevia{We illustrate the versatility of Virtuoso, by conducting five case studies that are \konrevia{time-consuming and difficult} to  accurately and rapidly \konrevic{assess} using existing simulation tools.}
    \item Virtuoso's source code and \konrevic{integration} with all five simulators is freely available at \textcolor{blue}{\url{https://github.com/CMU-SAFARI/Virtuoso}}.

\end{itemize}

\vspace{-3mm}
\section{Background \& Motivation}
\label{sec:motivation}

\textbf{VM Overheads.} 
Reducing the overheads of the VM subsystem is a long-standing challenge in computer architecture and OS research.
Lately, emerging data-intensive workloads\Workloads~ turned VM overheads into a major performance bottleneck.
As shown in multiple academic and industrial studies~\VMcharacterization, address translation can significantly degrade the performance of applications taking up to 40\% of the total execution time~\cite{contiguitas2023,radiantISMM21}. 
At the same time,  OS routines responsible for allocating physical memory  can cause high performance overheads, up to 95\%~\cite{hbdpISCA2020,minorfaultTACO2022,mementoMICRO2023}.

\begin{figure}[h!]
  \centering
  \includegraphics[width=\linewidth]{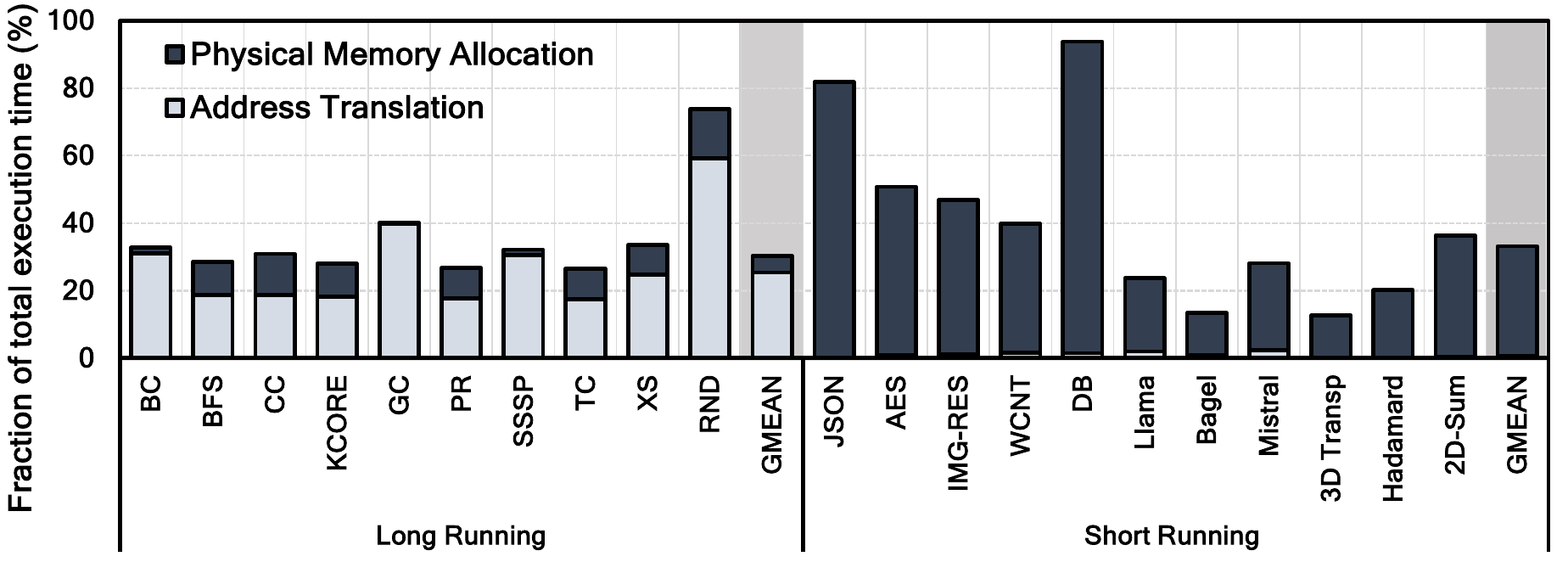}
  \caption{\konrevic{Fraction} of total execution time spent in address translation and physical memory allocation \konrevie{in long-running and short-running workloads} executed \konrevie{on a real high-end server system}~\cite{kratos20}.}
  \label{fig:motivation}
  \vspace{-3mm}
\end{figure}

Figure~\ref{fig:motivation} shows the portion of the total execution time spent on address translation and allocating physical memory
\footnote{We consider physical memory allocation as the total time spent in the \konrevic{page fault handler}. We populate the page cache before the application starts executing 
to demonstrate the overheads of the page fault handler even in the absence of long-latency major page faults (i.e., disk accesses).}
for long-running (\konrevic{i.e., > 100 s}) and short-running (\konrevic{i.e., < 1 s}) workloads executed in a real high-end server-grade system 
(our evaluation methodology is described in detail in \S\ref{sec:methodology}).
We make two key observations. First, long-running workloads spend on average 25\% \konrevic{(4.9\%)} of the total execution time on address translation \konrevic{(memory allocation)}. \konrevid{In contrast}, in short-running workloads the overheads of memory allocation take a large portion of the total execution time, i.e., 32\% on average, while the overheads of address translation are \konrevic{very small}, i.e., less than 1\% on average.
This is \konrevid{because} in long-running workloads, the overheads of physical memory allocation tend to be amortized over time, whereas in short-running workloads they are not. We conclude that the overheads of the VM subsystem can vary across different workloads and can heavily affect performance.

The increasingly data-intensive nature of emerging applications and the transition towards large physical address spaces~\cite{intelicelake} (e.g., via compute-enabled memory modules~\cite{ahnISCA2015,impicaICCD2016,ahn2015pim,ghose.ibmjrd19,mutlu2019processing,gao2016hrl}, large hybrid memory hierarchies\HybridNVM, memory disaggregation\MemoryCXL, heterogeneous systems with unified virtual memory\GPUUVM)
is expected to increase the overheads caused by the VM subsystem~\cite{guo2022clio,radiantISMM21}.

\textbf{Hardware/OS \konrevif{Co-Design.}} A promising way to alleviate the overheads of VM is to co-design the hardware and OS.
As shown in multiple prior works, VM can be improved via
(i) designing more efficient page tables~\cite{elastic-cuckoo-asplos20,hashJovanHPCA2023,hash_dont_cache,vm25,flataAsplos2022} (e.g., hash-based page tables~\cite{elastic-cuckoo-asplos20,hashJovanHPCA2023,hash_dont_cache}),
(\nb{ii}) enforcing and leveraging contiguity between virtual and physical addresses to increase the address translation reach of the processor\VMcontiguity~(e.g., range-based translation~\cite{karakostas2015}), 
(\nb{iii}) employing hash-based virtual-to-physical mappings to reduce the size of metadata used for address translation\VMrestrictive,  
(\nb{iv}) introducing intermediate address spaces\VMintermediate~to delay address translation until a main memory access,
(\nb{v}) employing large OS-managed TLBs~\cite{pomtlbISCA2017,ducati} to improve the TLB hit rate,
and (\nb{vi}) accelerating OS routines that manage the VM subsystem by offloading them to specialized hardware~\cite{hbdpISCA2020,minorfaultTACO2022,mementoMICRO2023}.

\begin{table*}[hb!]
  \centering
  \scriptsize

  \caption{Virtual memory schemes supported by existing simulators and Virtuoso (our proposed simulator).}
  \vspace{-2mm}
  \centering
  \begin{tblr}{
  colspec = {Q[0.4cm,c] | [1pt] Q[1.9cm,c] || [1pt] Q[3cm,c] | Q[2.2cm,c] | Q[2.2cm,c] | Q[1.8cm,c] | Q[1.6cm,c] | Q[1.35cm,c]}, 
    rowspec = {|c|c|c|c|c|c|c|},
    hline{1,Z} = {1pt},
    hline{2} = {1pt},
    hline{2} = {1pt},
    hline{3-6} = {0.5pt}, 
    hline{6-9} = {0.5pt},
    hline{10-18} = {0.5pt},
    vlines
  }
  \textit{Type} & \textbf{Simulator/\\ Component} & {\textbf{TLB}\\\textbf{Hierarchy}} & \textbf{Page Table \\Design} & \textbf{Contiguity\\ Schemes} & {\textbf{Intermediate}\\\textbf{Address Space}} & \textbf{Hash-based\\ Translation} & {\konrevic{\textbf{Memory \\Tagging}}} \\ 
  \hline
  \SetCell[r=8]{m} \rotatebox{90}{\makecell{\textbf{Emulation-based}}} & \textbf{SimpleScalar}~\cite{simplescalar}  & Generic TLB Controller &  \xmark &   \xmark &  \xmark & \xmark & \xmark   \\
  & \textbf{Multi2Sim}~\cite{multi2sim} & Generic TLB Controller &  \xmark &   \xmark &  \xmark & \xmark & \xmark   \\
  & \textbf{Scarab}~\cite{scarab} & \xmark & \xmark & \xmark & \xmark & \xmark & \xmark  \\
  & \textbf{Ramulator2}~\cite{luo2023ramulator2}  & \xmark & \xmark & \xmark & \xmark & \xmark & \xmark  \\
  & \textbf{ZSim}~\cite{sanchez2013zsim} & \xmark & \xmark & \xmark & \xmark & \xmark & \xmark  \\
  &\textbf{gem5-SE}~\cite{gem5} & {Generic TLB Controller} & {x86-64 \& ARM  PT} & \xmark & \xmark & \xmark & \xmark    \\
  &\textbf{ChampSim}~\cite{champsim} & {Generic \& TLB Prefetching}  &  {x86-64 PT } & \xmark & \xmark &  \xmark & \xmark  \\   
  &\textbf{Sniper}~\cite{sniper} & {Generic TLB Controller} & {Fixed PTW latency} & \xmark & \xmark& \xmark& \xmark  \\
  \hline
  \SetCell[r=3]{m} \rotatebox{90}{\makecell{\textbf{Full} \\ \textbf{System}}} & \textbf{PTLsim}~\cite{ptlsim} & Generic TLB Controller & {x86-64 \& ARM PT} & \konrevic{Linux THP~\cite{corbet2011,corbet2017}}  & \xmark & \xmark & \xmark    \\
  &\textbf{QFlex}~\cite{qflexv2.0} & Generic TLB Controller & {x86-64 \& ARM PT} & \konrevic{Linux THP~\cite{corbet2011,corbet2017}}  & \xmark & \xmark & \xmark \\
  &\textbf{Gem5-FS}~\cite{gem5} & Generic TLB Controller & {x86-64 \& ARM PT} &  \konrevic{Linux THP~\cite{corbet2011,corbet2017}}  & \xmark & \xmark & \xmark  \\
  \hline
  \SetCell[r=6]{m} \rotatebox{90}{\makecell{\textbf{Imitation}-\textbf{based}}}&  \SetCell[r=6]{m} \textbf{Virtuoso \\(this work)} & {Configurable TLB hierarchy}  & \SetCell[r=2]{m} {Hash-based PTs: ECH~\cite{elastic-cuckoo-asplos20},~HDC~\cite{hash_dont_cache}} & \SetCell[r=2]{m} Direct Segments~\cite{directsegments}  & \SetCell[r=3]{m}  Midgard ~\cite{midgard} & \SetCell[r=3]{m} Hash-based translation \newline \cite{nearmemoryPact17} & \SetCell[r=3]{m} Mondrian Data Protection\newline\cite{mondrian}  \\
  &  & {Multi-page size TLBs}  &                                                         &   &   &  &   \\
  &  & {Page-size prediction~\cite{papadopoulou2015}}  & \SetCell[r=2]{m}{Configurable Radix-PT + PWCs~\cite{isca2010-barr-trancache}}    &   \SetCell[r=2]{m} {Range Translation \& Eager Paging~\cite{karakostas2015}}  &  &  &   \\
  &  & {TLB prefetching~\cite{vavouliotis2021}} &  & ~\cite{corbet2011}  & \SetCell[r=3]{m} Virtual Block Interface~\cite{vbi} & \SetCell[r=3]{m} \konrevic{Hybrid} Restrictive \& Flexible Physical Segments~\cite{kanellopoulos2023utopia} & \SetCell[r=3]{m} Expressive Memory \newline\cite{xmem}   \\
  &  & {Software-managed TLBs~\cite{pomtlbISCA2017}}  & \SetCell[r=2]{m}{Support for nested TLB~\cite{nestedTLBNinjaICCD2024} and PTW~\cite{amdnested}}  & \SetCell[r=2]{m} {Linux-like~\cite{corbet2011,corbet2017} \& Reservation-based THP~\cite{reserve}}  &  &  &   \\
  &  & {TLB entries stored \newline in data caches~\cite{kanellopoulosMICRO2023victima}}  &   &   &  &  &   \\ 
  \end{tblr}
   \label{tab:simulators}
  \end{table*}

\textbf{Need for Detailed Simulation.} Given the \konrevic{large} VM overheads, it is critical to have methods for easily and quickly \konrevid{prototyping and evaluating} existing and new VM \konrevid{ideas and techniques}.
However, such an evaluation is challenging since VM components (i) span across the hardware/software boundary\nb{,} and (ii) are highly interdependent, which leads to 
significant variability in the overheads of the VM components across different workloads and system states.
For example, the effectiveness of TLBs~\cite{vm6,hybridtlbISCA2017} as well as the storage requirements, lookup latency and main memory contention caused by the page table heavily depend on the \konrevic{number} of large pages (e.g., 2MB pages) \nb{that} the OS's physical memory allocator provides \konrevic{to user applications}. At the same time, the physical memory \konrevic{allocation} policy affects the latency of the page fault handler which might heavily affect the tail latency of the application. 
Therefore, it is challenging to accurately model the overheads of the VM components with simple first-order models (e.g., \konrevid{those} that assume a fixed latency). 
We use two example cases to showcase the variability in the overheads caused by the VM components.

\begin{figure}[h!]
  \centering
  \includegraphics[width=1.0\linewidth]{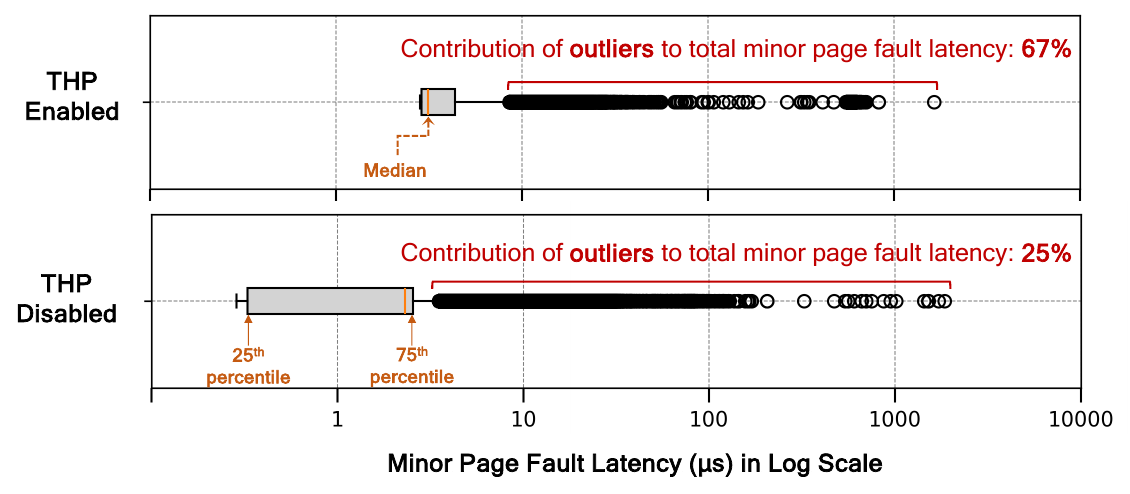}
  \vspace{-6mm}
  \caption{\konrevie{Minor page fault latency distribution across two different physical memory allocation policies (i.e., THP~\cite{corbet2011,corbet2017} enabled and disabled) measured in a real system~\cite{kratos20}.}}
  \label{fig:variation_mpf}
    \vspace{-4mm}
\end{figure}

\textbf{Example: Variation of Minor Page Fault Latency.} Fig.~\ref{fig:variation_mpf} shows the distribution of the minor page fault (MPF) latency using two OS page allocation policies, (i.e., transparent huge pages (THP)~\cite{corbet2011,corbet2017} enabled and disabled) across all workloads executed in a real high-end server-grade system (\S\ref{sec:methodology}).
We make two key observations. First, the latency of MPFs can vary significantly given a single physical memory allocation policy. With THP-enabled, 
the average MPF latency is \konrevid{$2.2\mu$s} while the standard deviation is larger than \konrevid{$50\mu$s}. 
Second, the distribution of the PF latency can significantly change when the physical memory allocation policy \konrevic{provides} large pages.
With THP-enabled, the contribution of the outliers (i.e., MPFs with latency larger than \konrevid{$10\mu$s}) to the total MPF latency is 67\%  while with THP-disabled, the contribution of the outliers to the total PF latency 25.5\%. \sepherd{Prior works (e.g., \konrevic{~\cite{cbmm,mansi2024characterizingphysicalmemoryfragmentation}}) attribute this variability to the large number of different operations (e.g., page zeroing, fallback mechanism, huge page allocation, page table updates, memory reclamation) and pathological cases that might occur during page fault handling.}

\begin{figure}[h!]
  \vspace{-1mm}
  \centering
  \includegraphics[width=1.0\linewidth]{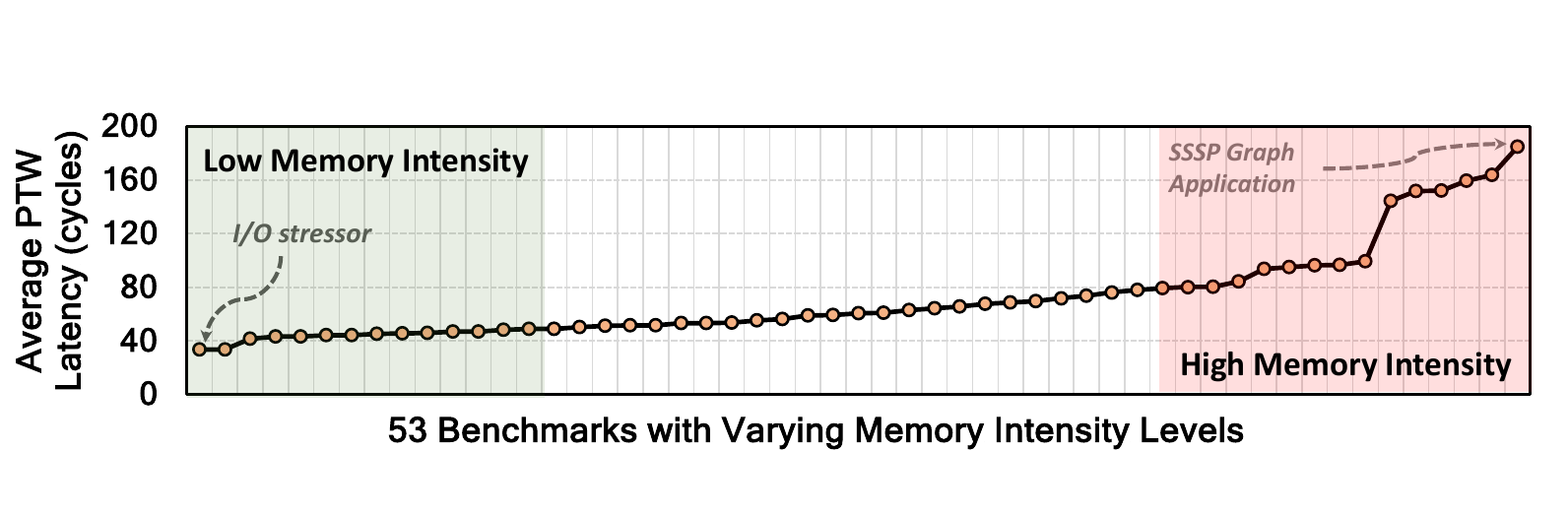}
  \vspace{-7mm}
  \caption{Average PTW latency across \konrevic{53 different applications that exhibit varying levels of memory intensity, measured in a real \konrevid{high-end server} system~\cite{kratos20}.}}
  \label{fig:variation_ptw}
  \vspace{-2mm}

\end{figure}

\textbf{Example: Variation of Page Table Walk (PTW) Latency.} Fig.~\ref{fig:variation_ptw} shows the average PTW latency 
across 45 applications executed in a real system that stress VM at different levels\footnote{We \konrevic{use} different configurations of \konrevic{the} \texttt{stress-ng} benchmarks~\cite{stress-ng} and the long-running workloads described in \S\ref{sec:methodology}. We measure the page table walk latency using performance counters.} 
We observe that the PTW latency \konrevic{significantly varies} across different applications. 
For example, the PTW of an application that performs large \texttt{I/O} allocations
\konrevic{is 39 cycles while the PTW latency of the single-source shortest path workload (SSSP) from GraphBig~\cite{Lifeng2015} is larger than 180 cycles.}

We conclude that the overhead of the VM subsystem significantly varies across different workloads and system configurations and thus, cannot be accurately modeled with first-order approximations \konrevic{(e.g., assuming fixed latencies)} but requires detailed simulation.

\subsection{Existing Simulation Frameworks}

We classify \konrevic{existing} simulators (e.g., ~\cite{multi2sim,sanchez2013zsim,sniper,gem5,ramulator,simplescalar,ptlsim,qflexv2.0,scarab}) into two broad \konrevic{categories}: (i) simulators that \emph{emulate} OS routines, and (ii) \emph{full-system} simulators where a real full-blown OS is executed on top of a hardware simulator.
\konrevic{Unfortunately, as we describe below, neither type of simulator is well-suited} for evaluating VM schemes that rely on co-designing OS routines and hardware support,
which hinders fast and accurate protyping and evaluation of such schemes.
Table~\ref{tab:simulators} summarizes the VM components supported by \konrevif{eleven} existing simulators and by our proposed simulator, Virtuoso. %

\begingroup
\sloppy
\textbf{Emulating OS Routines.} 
Many existing simulators (e.g., \Emulationbased) are designed 
with a focus on accurately modeling the core, main memory or other hardware
components that do not directly rely on or interact with the OS. Hence, these simulators lack (and \konrevic{some} do not \konrevic{need for the use cases they are designed for}) a methodology to
accurately model the implications (e.g., latency, memory interference) of the OS components involved in the VM subsystem.
For example, multiple simulators \konrevic{(e.g., ~\cite{sanchez2013zsim,sniper,champsim})} model \emph{only the} functional interactions of the application with a subset of OS routines (e.g., \konrevic{\emph{mmap()}~\cite{mmap}}) and typically use first-order approximations (e.g. Sniper~\cite{sniper} uses a fixed PTW latency and Champsim~\cite{champsim} uses a fixed page fault latency) to model VM overheads. 
However, as we show in Fig.~\ref{fig:variation_mpf} and Fig.~\ref{fig:variation_ptw}, the overheads of VM can significantly vary across different workloads and applications, and hence, cannot be accurately modeled with static first-order approximations. In \S\ref{sec:validation}, we show that the baseline version of Sniper that uses a fixed PTW latency leads to ~35\% error in IPC estimation compared to the real system.
Thus, such simulators are not a good fit for evaluating new VM schemes that require changes to the OS kernel code and new hardware support.

\textbf{Full-System Simulation.}
Full system simulators (e.g., \Fullsystem) like the full-system execution mode provided by gem5~\cite{gem5} and QEMU-based architectural simulators like QFlex~\cite{qflexv2.0} enable the execution of a full-blown OS, including realistic memory management and other OS routines, on top of a hardware simulator. 
Such a methodology is particularly valuable when evaluating VM designs that involve changes to the OS kernel code and require new hardware support.
However, existing full-system simulation methodologies have three main limitations: (i) low simulation speed, (ii) high memory overheads, and (iii) high development time and effort. 
First, simulating a full-blown OS drastically increases simulation time and memory consumption, hindering rapid design space exploration. 
Simulating every single OS routine without the possibility of omitting those that are irrelevant to the desired evaluation can significantly increase simulation times. At the same time, spawning a full-blown OS significantly increases memory consumption per simulation task. 
In \S\ref{sec:simcomp}, we show that simulating a full-blown OS on top of \konrevic{gem5~\cite{gem5}} can increase simulation time by 77\% and memory consumption by \konrevic{1.69x} (from 1GB to 1.69GB per simulation task) compared to the system call emulation mode of gem5 \konrevid{(\konrevid{gem5-SE})}. Second, evaluating new hardware/OS co-design schemes \konrevic{on top of full-system simulators necessitates} (i) the modification of an already complex OS kernel code, 
(ii) its \konrevic{functional} verification of top of simulated hardware and (iii) simulator
extensions to support new hardware components \konrevic{(e.g., new TLB design\konrevid{s})}, and (iv) \konrevic{complex modifications to the interface} between the OS routines and the hardware. This process requires significant development effort and time, especially for researchers who are not experts in OS development. \konrevic{We conclude that, while full-system simulators are indispensable tools in computer architecture research, they limit productivity and cause high simulation overheads, thereby hindering their practical utility in \konrevid{exploring and evaluating} VM schemes that span across the hardware/software boundary.} 

\textbf{Simulation Requirements.} To evaluate new VM schemes accurately, efficiently and rapidly, a simulation framework needs to  (i) enable fast prototyping of the required hardware and OS modifications, (ii) accurately and quickly estimate 
the overheads caused \konrevic{and the benefits provided by} the new OS and hardware components, \konrevic{(iii) model the interaction of the VM components with the rest of the system and between each other}.
\endgroup

\section{Virtuoso: Overview}

We present Virtuoso, a new simulation framework that enables fast and accurate prototyping and evaluation of the \textit{software and hardware components} of the VM subsystem.
The key idea of Virtuoso is to employ a lightweight userspace kernel, written in a high level language (\konrevic{e.g.,} C++), that enables researchers to 
(i) isolate the functionality of \emph{only the desired kernel code} to speed up simulation time,
(ii) easily \emph{develop new OS routines} using a high-level language without being kernel code experts, and 
(iii) \emph{accurately} evaluate the application- and system-level implications of the OS by integrating Virtuoso into an architectural simulator.

\begin{figure}[h!]
    \centering
    \includegraphics[width=1.0\linewidth]{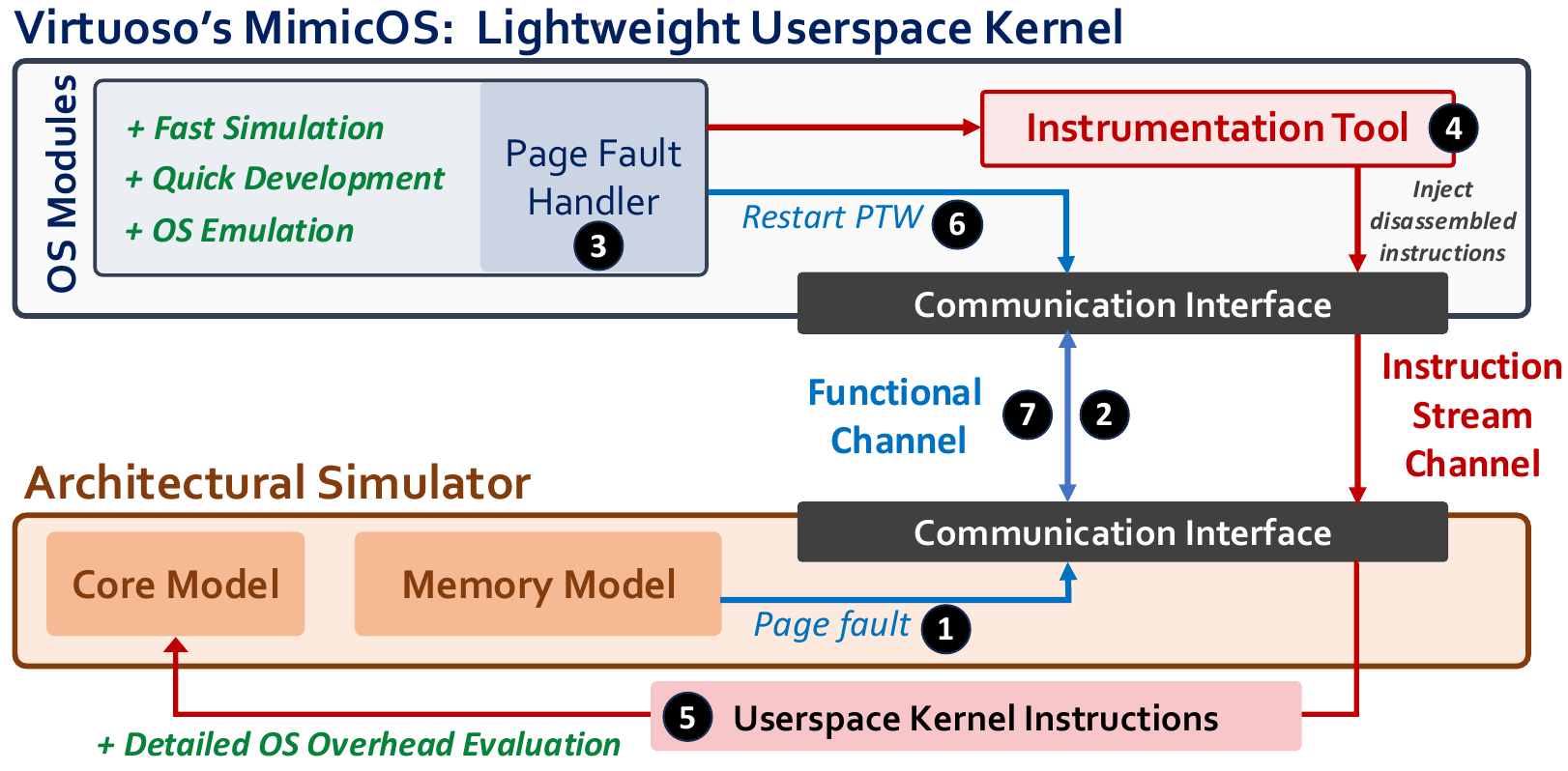}
    \caption{\konrevic{Overview of Virtuoso's Architecture.}} 
    \label{fig:virtuoso_overview}
    \vspace{-2mm}
\end{figure}

Figure \ref{fig:virtuoso_overview} illustrates a high-level overview of Virtuoso's components and workflow.
Virtuoso consists of two main components: (i) a lightweight userspace kernel, called MimicOS, that imitates the virtual memory subsystem of the OS,
and (ii) a communication channel between MimicOS and the architectural simulator \konrevic{that Virtuoso is coupled with}.
When the architectural simulator executes an event that requires OS intervention (e.g., page fault, memory allocation, etc.)~\circled{1}, 
the simulator forwards the event to MimicOS through the communication channel~\circled{2}.
MimicOS processes the event~\circled{3} and Virtuoso performs two operations. First, Virtuoso dynamically instruments MimicOS's binary~\circled{4} and
injects MimicOS's \konrevic{disassembled instructions} into the processor performance model of the simulator~\circled{5}.
This way, the simulator can accurately estimate the performance implications of the executed OS routines on the application.
Second, when MimicOS resolves the event, it returns the functional response to the architectural simulator \konrevid{(e.g., signals the core to restart walking the page table~\circled{6}) through the functional  channel~\circled{7}.}

\section{Imitation-\konrevid{B}ased Simulation Methodology}
\label{sec:imitation}

\konrevic{We describe} the key components of Virtuoso's simulation methodology, (i) the lightweight \konrevic{userspace} kernel and
(ii) the communication interface between the kernel and the architectural simulator, and 
provide a step-by-step example of the simulation flow of a page fault handling routine.

\subsection{Lightweight \konrevic{Userspace} Kernel}

Virtuoso employs a lightweight userspace kernel to imitate the functionality of the desired OS kernel code. 
Such a design decision enables researchers to 
(i) simulate only the relevant OS routines to speed up simulation time, and 
(ii) quickly and easily develop new OS modules.

 \head{Kernel Module Selection} 
 Virtuoso's kernel comprises different modules selected by the researcher to balance accuracy and simulation time depending on their research needs. 
For example, a kernel may \emph{solely} comprise of a page fault handler \konrevic{if}
the researcher wants to quickly evaluate the impact of different page fault handling mechanisms \konrevic{on} system performance without 
taking irrelevant OS routines (e.g., thread scheduler) into consideration. 
As we demonstrate in \S\ref{sec:simcomp}, executing a simulator paired with a userspace kernel that faithfully mimics the functionality 
of \emph{only} the \ak{Linux} memory management subsystem, is 49\% faster than simulating the \konrevic{entire} Linux kernel.

\head{Ease of Development} 
The \konrevic{userspace} kernel can be written in a high-level language (e.g., Python, C++), which enables easier development of new OS routines without requiring expert knowledge. 
For example, the researcher can easily develop a new machine learning-based page replacement algorithm using a high-level library 
(e.g., \konrevid{m}lpack~\cite{mlpack2023}, \konrevic{T}ensor\konrevic{F}low~\cite{tensorflow}, \konrevic{P}y\konrevic{T}orch~\cite{torch}) and integrate it with the kernel without needing to understand or modify the complex code of a production-grade OS. 
At the same time, Virtuoso's modular design allows increasing the number of supported OS modules to closely mimic the functionality of a target kernel at the cost of increased simulation time.

\subsection{Interface with the Architectural Simulator}

To evaluate the impact of OS routines on the performance of \konrevic{a} system, the userspace kernel needs to execute on top of an architectural simulator.
To achieve this, Virtuoso (i) executes both processes (i.e., the userspace kernel and the simulator) as standalone applications and (ii)
establishes a new communication interface between the userspace kernel and the simulator that consists of 
two new communication channels that employ synchronization primitives to orchestrate the execution flow between the kernel and the simulator.

\head{Communication Channels}
Virtuoso establishes two communication channels between the kernel and the simulator: (i) a \emph{functional} and (ii) an \emph{instruction stream} channel.
Through the functional channel, the simulator communicates functional requests (e.g., page fault \ak{requests}) to the kernel and the kernel communicates the emulated result of the request back to the simulator (e.g., signal to restart the page table walk).
However, the functional channel is not sufficient to estimate the impact of the OS routines on the performance of the system.
For example, the architectural simulator cannot estimate the impact of the page fault handler \konrevic{on various system components} (e.g., main memory \konrevic{controller} contention) by using only
the functional state (e.g., the physical address of the new page) of the userspace kernel.
To address this issue, \nb{Virtuoso executes the userspace \konrevic{kernel} \konrevic{a} binary instrumentation tool (e.g., Intel \konrevic{Pin}~\cite{intel_pin}, DynamoRIO~\cite{dynamorio}) to dynamically generate the kernel's instruction stream (e.g., the page fault handler instructions) and communicates 
it to the simulator through a separate \emph{instruction stream channel}.}

\head{Synchronization Primitives}
To achieve high simulation speed while maximizing portability (i.e., porting the userspace kernel to many \konrevic{different} architectural simulators with minimal changes), Virtuoso employs (i) POSIX-based~\cite{posix} shared memory primitives to exchange 
messages between the kernel and the architectural simulator, and (ii) \emph{magic} operations (e.g., \texttt{m5ops} in gem5~\cite{gem5}, \texttt{xchg} instructions in Sniper~\cite{sniper}) to synchronize the execution of the userspace kernel
with the architectural simulator.\footnote{\sepherd{Magic operations are special instructions
that may or not be part of the ISA and are used to notify the simulator to perform a specific action. For example, when Sniper~\cite{sniper} decodes the \texttt{xchg R1,R2} instruction, and r1 is identical to r2, it treats it as a signal 
to perform a specific special action dictated by the content of r1 (e.g, start detailed simulation).}}

\head{Execution Flow}
When the simulated application causes an interrupt or a system call, the architectural simulator performs two actions: (i) writes the interrupt/system call parameters to \konrevic{the functional channel (i.e., a \konrevie{POSIX-based} shared memory segment~\cite{shmem})} and (ii) notifies the userspace kernel to read the parameters and start processing the request.
While the userspace kernel processes the request, the binary instrumentation tool produces the instruction stream of the kernel's code and sends it to the simulator through the instruction stream channel. The simulator consumes the instruction stream, feeds it to \konrevic{its} core model, and estimates the impact of the kernel's code on performance. The production and the consumption of the kernel's instruction stream happen in parallel to avoid 
unnecessary latency in the simulation.\footnote{The latency for the production of the kernel's instruction stream could be hidden by using a runahead thread~\cite{runahead-mutlu-2003,ramirez2008runahead}. Such an optimization is useful especially when the simulator's frontend is trace-based and all the instructions of the application are known in advance.}
When the userspace kernel resolves the request, it performs two actions: (i) writes the result of the request to the \konrevic{functional channel} and (ii) executes a magic instruction to signal the simulator to continue the simulation of the application. 
When the simulator decodes the magic instruction, it pauses the instrumentation of userspace kernel instructions and switches back the simulated application.

\begin{figure*}[ht!]
  \centering
  \includegraphics[width=0.9\linewidth]{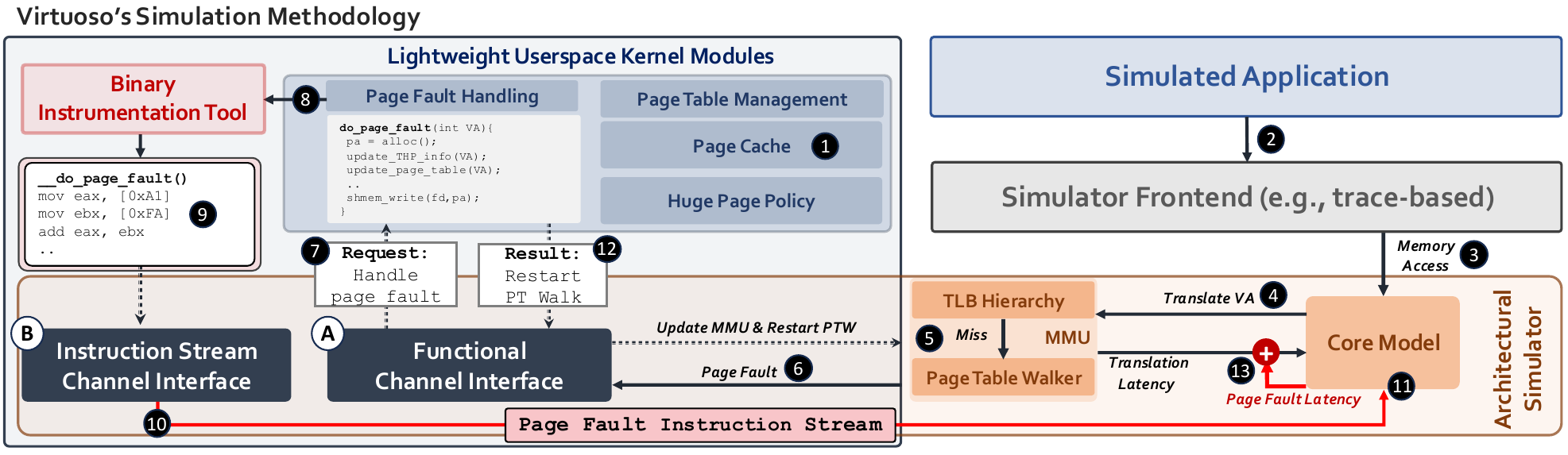}
  \caption{Example page fault handling workflow of Virtuoso \konrevic{coupled with} an architectural simulator.}
  \label{fig:virtuoso_workflow}
\end{figure*}

\subsection{\konrevic{Multithreaded} Userspace Kernel}

Virtuoso's userspace kernel supports multithreading to concurrently handle multiple system calls or interrupts from different processes.
To achieve this, when an application being executed on the simulator issues a request to the kernel, the kernel spawns a new thread to handle the request or forwards the request to an available thread. 
The kernel uses synchronization primitives to guarantee the correctness of the kernel routines in \konrevic{multithreaded} environments and model the performance overheads 
of atomic operations. For example, if multiple applications compete for physical memory resources, our methodology can capture the corresponding synchronization overheads.

\subsection{Simulation Flow:  Page Fault Handling Example}
Figure~\ref{fig:virtuoso_workflow} demonstrates the workflow of the proposed simulation methodology with an example case study of a page fault (PF) handler.
First, the kernel and the simulator are launched as userspace processes.
In this example, the kernel comprises a PF handler with multiple different modules ~\circled{1} (e.g., page table management, page cache~\cite{pagecache} management\konrevic{,} etc.).
The simulated application is fed to the frontend (i.e., instruction format generator) of the simulator (e.g., trace-based, instrumentation-based, emulation-based etc.) to generate the instruction stream~\circled{2}.
\konrevic{If an instruction contains a load or store memory operand, the frontend issues a memory access request to the core model of the simulator~\circled{3}.}
The core model forwards the memory request to the memory management unit (MMU) model to perform address translation~\circled{4}.
If the MMU does not find the translation in the TLB hierarchy, it triggers a page table (PT) walk~\circled{5}.
In this scenario, the PT walker does not find the translation in the PT and triggers a PF~\circled{6}. 
Through the functional channel~\circledwhite{A}, the simulator sends a request to the kernel to handle the PF~\circled{7}.
The kernel decodes the message and executes the PF handler code~\circled{8}.
The PF handler code is instrumented using a binary instrumentation tool (e.g., Intel \konrevic{Pin}~\cite{intel_pin}, DynamoRIO~\cite{dynamorio})~\circled{9} and the instrumented disassembled instruction stream is sent to the simulator through the instruction stream channel~\circledwhite{B}.

The PF \konrevid{handler's} instruction stream is forwarded~\circled{10} to the core model of the simulator \ak{and the simulator} \konrevid{models the execution of} the kernel's instructions to estimate the impact of the PF handler on the microarchitectural state and performance (e.g., main memory contention, cache pollution)~\circled{11}. 
When the PF handler completes executing, the kernel communicates the outcome of the PF (e.g., the physical address of the new page and the page size) to the simulator~\circled{12}.
The simulator then re-walks the PT, the core model adds the latency of the PF to the translation latency~\circled{13} and forwards the physical address to the memory hierarchy.

\section{MimicOS: A Lightweight Userspace Kernel for Memory Management}

Using \konrevic{our new} imitation-based simulation methodology (\S\ref{sec:imitation}), we build MimicOS, a new lightweight kernel written in C++ that mimics, 
but is not limited to, the basic memory management functionality of the Linux kernel~\cite{linuxkernel} \konrevic{for x86-64 systems~\cite{intelx86manual}}.

\subsection{\konrevid{Mimicking} Linux Memory Management }
\label{sec:MimicOS}

\begin{figure}[ht!]
    \centering
    \includegraphics[width=1.0\linewidth]{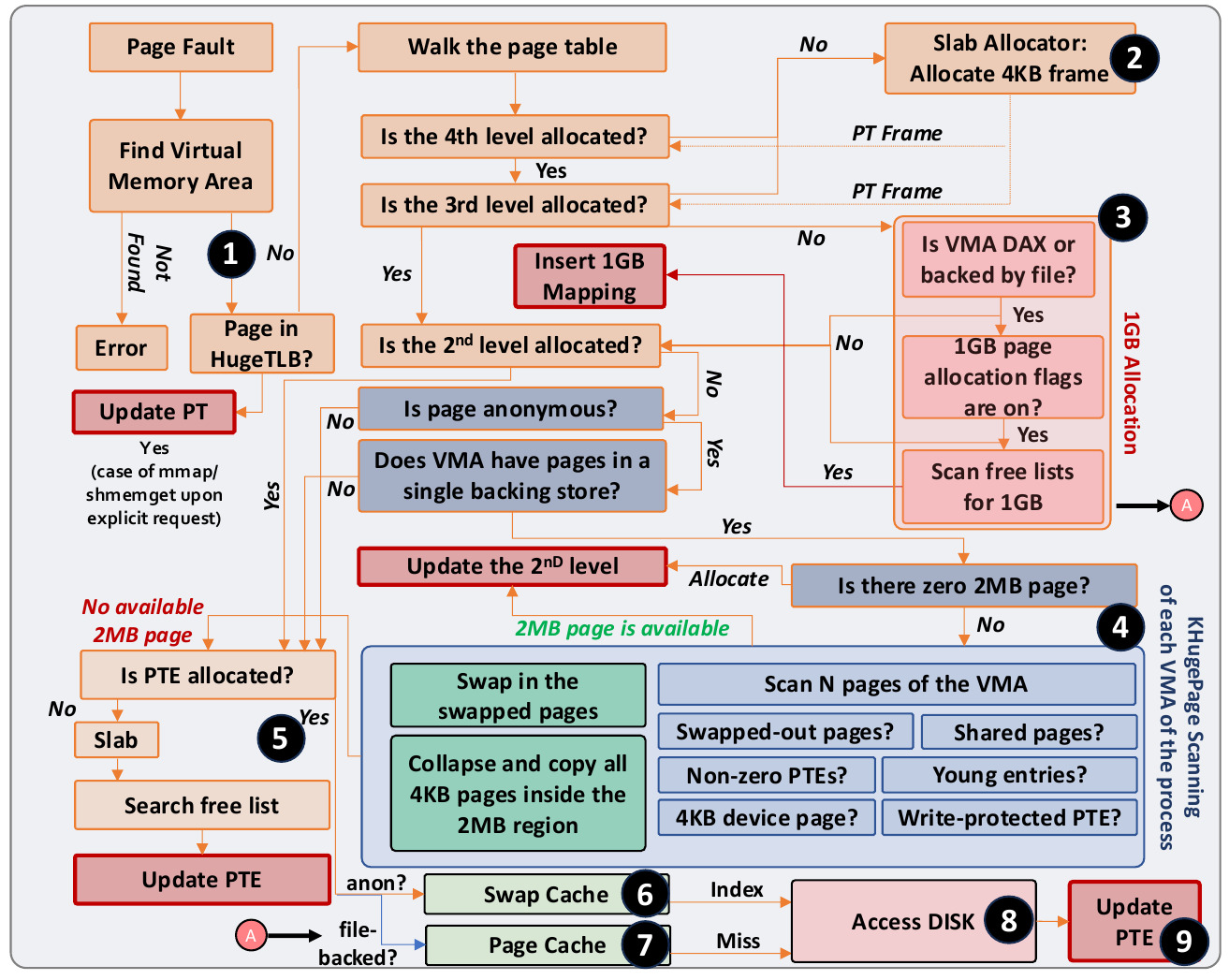}
    \caption{\konrevc{MimicOS Memory Management Subsystem.}}
    \label{fig:mimicos_loc}
    \vspace{-2mm}
  \end{figure}

As shown in Fig.~\ref{fig:mimicos_loc}, MimicOS employs a memory management scheme that mimics the one used by Linux. On a page fault, MimicOS checks if the virtual memory area (VMA)~\cite{vma} should be stored in hugetlbfs\footnote{\konrevic{\textit{hugetlbfs}~\cite{hugetlbfs} is a Linux kernel policy responsible for reserving huge pages to ensure availability during allocation time. A virtual memory area is mapped through \textit{hugetlbfs} only when large pages are explicitly requested via mmap() or shmemget() calls.}}~\cite{hugetlbfs}~\circled{1} and updates the page table (PT). If not, MimicOS begins walking the PT. To allocate new PT frames (in case of a page fault), MimicOS requests new frames from the slab allocator~\cite{slab}~\circled{2}.
If the 3rd-level PT entry is uninstantiated, MimicOS decides whether \konrevic{or not} to allocate a 1GB physical page based on three conditions~\circled{3}: (1) the VMA uses DAX~\cite{daxvm} or is backed by a file, (2) 1GB allocation flags are set, and (3) a 1GB contiguous physical memory region is available in the buddy allocator’s free list. If all conditions are met, a 1GB page is allocated, data is fetched from the page cache (or disk), and the PT is updated. If not, MimicOS attempts to allocate smaller pages and resumes the PT walk.
For empty 2nd-level PT entries, MimicOS attempts allocating a 2MB page if the VMA is anonymous~\cite{vma}~\circled{4}. If a \konrevid{zeroed} 2MB page is available, MimicOS allocates it, and updates the PT. If not, a 4KB page is allocated, the final PT level updated~\circled{5}, and \texttt{khugepaged}~\cite{khugepage} is notified to scan memory and merge 4KB pages into 2MB pages. 
\konrevic{If the PTE is allocated and corresponds to anonymous pages, MimicOS accesses the swap cache~\cite{swapcache} to retrieve the location of the data in the swap file~\cite{swapspace}~\circled{6}.
If the PTE is empty and corresponds to file-backed pages (e.g., data originates from files), MimicOS accesses the page cache~\cite{pagecache} (software data structure that resides in memory and stores recently-accessed file-backed pages) to retrieve the data~\circled{7}.
On a page cache miss or swap access, MimicOS fetches the data from disk (we simulate the disk access latency using an SSD simulator~\cite{tavakkol2018mqsim})~\circled{8} and updates the PT~\circled{9}.}

\subsection{VirTool: A \konrevic{Toolset} for VM Research}

\konrevic{We} integrated MimicOS with (i) four architectural simulators: Sniper~\cite{sniper}, Ramulator~\cite{ramulator}, ChampSim~\cite{champsim}, 
and gem5-SE~\cite{gem5}, and (ii) an SSD simulator, MQSim~\cite{tavakkol2018mqsim}, 
to enable the evaluation of storage device impact on VM.
By doing so, we aim to unlock a wide range of new \konrevid{ideas and} case studies ranging from low-level microarchitectural VM schemes to \konrevic{hardware/software/OS co-design VM solutions.}
Using MimicOS+Sniper as a baseline, we create \emph{VirTool}, a comprehensive toolset of state-of-the-art 
VM~\cite{sniper}. Table~\ref{tab:simulators} provides an overview of the techniques included in VirTool. 
With VirTool we aim to provide a common ground for researchers to easily and \konrevic{consistently} develop and evaluate existing and new VM techniques.

\section{\nb{Extending Virtuoso}}

\subsection{Support for Virtualized Environments}
Virtuoso \konrevic{supports out-of-the-box simulation of virtualized execution environments} (i.e., virtual machines running on top of a hypervisor \konrevid{(e.g., \cite{goldberg1974survey,amdnested})}).
To achieve this, Virtuoso spawns two userspace kernels (MimicOSes): \nb{1)} one that acts and mimics the hypervisor (\konrevic{e.g.,} acting like KVM~\cite{KVM}) and 2) one that imitates the guest OS \konrevid{(e.g., Linux)}.
When the guest OS needs to send requests to the hypervisor, the same process described in \S\ref{sec:MimicOS} \konrevid{is} followed in a nested manner, so that the simulator captures the instruction stream of \konrevic{\emph{both}} the guest OS and the hypervisor.
VirTool already provides support for \konrevic{\emph{nested address translation}} \konrevic{\cite{amdnested}}, which is a key feature for modeling virtualized environments.

\subsection{Integration with Architectural Simulators}
\label{sec:integration-other}

At a high level, integrating Virtuoso with an architectural simulator \akdel{mainly} requires three key steps: %
(i) using an emulation, instrumentation or other tools (e.g., custom tracer) to capture the instruction stream generated by MimicOS and convert it to the format used by the architectural simulator,  
(ii) establishing a \konrevic{bi-directional} communication channel \konrevic{(e.g., POSIX-based shared memory~\cite{shmem})} \konrevic{between} MimicOS and the memory model (e.g., MMU model) of the architectural simulator to exchange messages \konrevic{(e.g., signal\konrevid{s} for interrupt, system call output)},
(iii) establishing a communication channel between MimicOS and the core model of the architectural simulator to inject the instruction stream generated by MimicOS. 
\konrevic{We already} integrated Virtuoso with  \konrevic{five} different simulators: Sniper~\cite{sniper}, Ramulator\konrevic{~\cite{ramulator, luo2023ramulator2}}, ChampSim~\cite{champsim},  gem5-SE~\cite{gem5} and MQSim 
~\cite{tavakkol2018mqsim}.

Table \ref{tab:virtuoso_integration} shows the \konrevic{additional} lines-of-code required for the integration.

\begin{table}[t!]
    \centering
    \footnotesize
    \begin{tabular}{| l | !{\vrule width 1.0pt} c|c|c|c|}
    \hline
    \texttt{Simulator} & \texttt{Frontend} & \texttt{Core model} & \texttt{MMU model} & \texttt{Files} \\
    \noalign{\hrule height 0.7pt}  %
    \hline
    \hline
    \textbf{ChampSim}~\cite{champsim} & 56 & 45 & 22 & 6 \\
    \textbf{Sniper}~\cite{sniper} & 46 & 35 & 180 & 9 \\
    \textbf{Ramulator2}~\cite{luo2023ramulator2} & 79 & 83 & 44 & 6 \\
    \textbf{{gem5-SE}}~\cite{gem5} & 0 & 221 & 44 & 12 \\
    \hline
    \end{tabular}
    \vspace{1em}
    \caption{
    \konrevic{Additional lines of code and number of files modified in different simulators to integrate Virtuoso.}}
    \label{tab:virtuoso_integration}
    \vspace{-2em}
\end{table}

\begin{figure}[h!]
    \centering
    \includegraphics[width=1.0\linewidth]{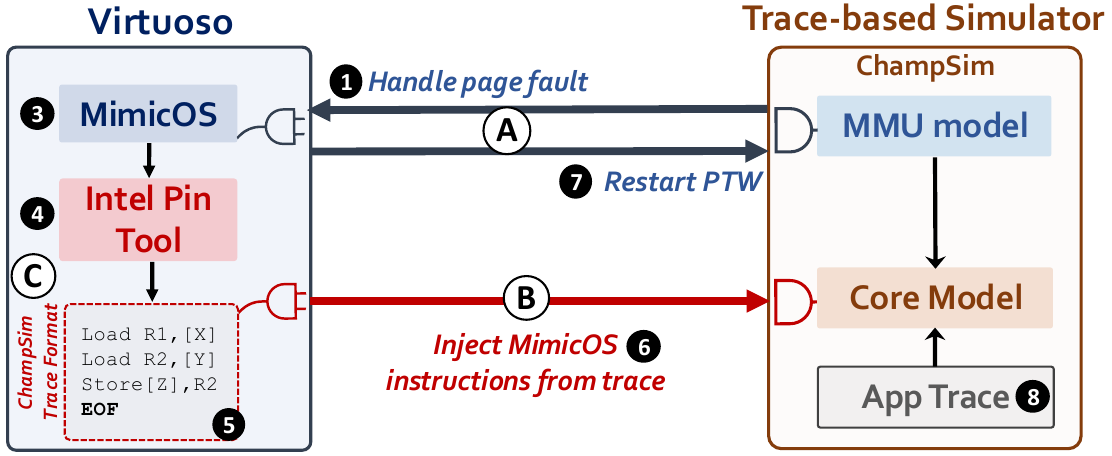}
    \caption{Integrating Virtuoso with trace-based simulators.}
    \label{fig:champsim_integration}
\end{figure}

\textbf{Simulators with Trace-based Frontend.} Trace-based simulators \konrevic{(e.g., ~\cite{champsim,ramulator,sniper,luo2023ramulator2,tavakkol2018mqsim,dramsim2})} typically \nb{simulate workloads using} input \nb{trace} files that represent the instructions and memory accesses of the workload \nb{generated by instrumentation and emulation tools (e.g., Intel Pin~\cite{intel_pin}) or other simulators.}
Virtuoso can be seamlessly integrated with trace-based simulators \nb{by following the steps described} in Fig.~\ref{fig:champsim_integration}. We use ChampSim~\cite{champsim} as an example trace-based simulator. First, MimicOS is booted in parallel with ChampSim and runs as a separate process on top of a binary instrumentation tool. ChampSim is modified in two ways: (i)
the MMU model \akdel{gets attached} to MimicOS using a bi-directional communication channel to receive and send functional requests~\circledwhite{A} and (ii) the core model \akdel{gets attached} to a communication channel to receive MimicOS's disassembled instruction stream~\circledwhite{B}. 
When the MMU model encounters a page fault, it sends a functional request to MimicOS to handle it~\circled{1}.
MimicOS starts executing the corresponding handler~\circled{3} and the binary instrumentation tool (e.g., Intel Pin~\cite{intel_pin}) generates the disassembled instruction stream~\circled{4}.
The instrumentation tool is modified to generate a trace that follows
the format expected by ChampSim~\circledwhite{C}. \konrevid{The instructions from MimicOS's trace~\circled{5} are streamed through the communication channel to
ChampSim's core model~\circled{6}, which models their execution.}
When the page fault is resolved, MimicOS notifies the MMU to re-walk the page table~\circled{7} and ChampSim's core model starts fetching instructions from the original application trace~\circled{8}. 

\sepherd{
\textbf{Simulators with Execution-driven Frontend.}
\konrevic{Execu\-tion-driven simulators}, such as Sniper~\cite{sniper}, Scarab~\cite{scarab} and ZSim~\cite{sanchez2013zsim},  dynamically instrument~\cite{intel_pin,dynamorio} the simulated application and generate the instruction stream on-the-fly without storing a trace file. Such a simulation methodology is particularly useful when the simulator manipulates the functional model (e.g., simulation of wrong path execution\konrevic{~\cite{gem5,scarab,eyerman2023simulating,mutlu2005analysis}}).
Virtuoso can be integrated with these simulators the same way as trace-based simulators with one key difference: when the instrumentation tool generates MimicOS's instruction stream, it directly injects it into the core model of the simulator without 
the need for an additional trace file. In this scenario, the core model of the simulator must be modified to dynamically switch between the instruction stream generated by MimicOS and the original instruction stream of the workload.}

\textbf{Simulators with Emulation-based Frontend.}
\sepherd{\konrevic{Simulators with \akdel{an} emulation-based frontend} (e.g., gem5~\cite{gem5}, QFlex~\cite{qflexv2.0}) use an emulation tool to capture the instruction stream of \akdel{the} workload and then feed the instructions to the core model of the simulator.
Integrating Virtuoso with these simulators is straightforward, as the existing emulation tool can be \konrevic{reused} to capture the instruction stream generated by MimicOS and feed it to the core model of the simulator.
For example, in Virtuoso's integration with gem5, when the MMU model encounters a page fault, it sends a request to MimicOS through shared memory \konrevic{and} the emulation tool produces the instruction stream of MimicOS, feeding it to the core model of gem5.} %

\subsection{Usage in Heterogeneous System Simulation}
\sepherd{
Virtuoso can be used to facilitate VM research in heterogeneous systems comprising of accelerators managed by a host CPU. One such example could be Unified Virtual Memory (UVM)\GPUUVM~that enables the use of a shared virtual address space across GPUs and CPUs. 
UVM management operations are typically orchestrated by the device driver running on the CPU (Host), using an Input-Output Memory Management Unit (IOMMU)~\cite{iommu}. 
In this scenario, Virtuoso's imitation-based methodology can be applied to model (i) functionalities provided by the OS and the device driver (e.g. host/device memory allocation, page migration) and (ii) functionalities of the IOMMU (e.g. page translation). 
Existing UVM-enabled GPU simulators~\cite{ganguly2019interplay,sun2019mgpusim} emulate events (e.g., page allocation, migration and translation) using fixed latencies or analytical models.
Consequently, integrating Virtuoso into such simulators requires 
(1) extending MimicOS to imitate the desired OS components (e.g., UVM driver\konrevic{~\cite{uvmdriver}}) and (2) establishing a communication channel between the host CPU simulator and the accelerator simulator to communicate the corresponding OS-related latency overheads.}

\sepherd{\subsection{Current Limitations} 
We believe that Virtuoso is a good fit for studies focusing on \konrevic{VM, which spans across the hardware and OS layers of the system stack.}
Virtuoso's speed and accuracy in simulating the Linux memory subsystem \konrevic{and} hardware MMU makes it particularly useful for academic research, system optimization, and the preliminary testing of hardware/OS changes before deployment on actual systems.
At the same time, researchers can expand MimicOS to incorporate more advanced OS functionality and adjust the accuracy \konrevic{and} simulation time as per their research requirements.
Hence, even though it provides a viable alternative to full-system simulators, we do not suggest that Virtuoso replaces them but rather complements them. 
In many cases, researchers need to simulate the entire system stack, including a real OS, to discover previously unknown performance bottlenecks or to evaluate the performance of a new hardware/OS cooperative technique in production-level OSes.
In such cases, full-system simulators like gem5~\cite{gem5} can provide a more accurate simulation of the entire system stack compared to Virtuoso. 
As Virtuoso evolves, further development could expand its capabilities, potentially bridging some of its current gaps \konrevic{with full-system simulators} and enabling \konrevic{the modeling of} more complex OS-level operations.}

\section{Virtuoso: Validation \& Use Cases}

\konrevic{We} (i) validate Virtuoso's accuracy against a real high-end server-grade CPU, (ii) evaluate Virtuoso's simulation time overheads when integrated into four different architectural simulators, and (iii) we conduct five diverse case studies to demonstrate Virtuoso's versatility.

\subsection{Evaluation Methodology}
\label{sec:methodology}
\textbf{System Configuration.}
We use the version of Virtuoso \nb{integrated with} Sniper~\cite{sniper} as our primary simulation tool. We chose Sniper for four key reasons: (1) it provides a good balance between microarchitecture, cache hierarchy, interconnect, main memory modeling details \konrevic{(we heavily refactored and enhanced the baseline DRAM model inspired from Ramulator~\cite{ramulator,luo2023ramulator2})} and simulation speed; (2) it is scalable in multi-core system simulation; (3) it is more programmer-friendly than gem5 ~\cite{gem5}; and (4) it achieves higher IPC performance estimation accuracy over gem5-SE~\cite{gem5}, as shown in prior studies~\cite{Akram201686CA} and as we also verified. 
Table~\ref{tab:simconfig} shows the configuration of the baseline simulated system, the configurations of all the schemes we evaluated in our case studies (\S\ref{sec:usecase1}-\ref{sec:usecase5}) and the configuration of the real system we validated Virtuoso against. 
Virtuoso along with all scripts, benchmarks, integration with \konrevic{five} simulators and all techniques included in VirTool, \konrevic{is freely} available at \url{https://github.com/CMU-SAFARI/Virtuoso}.

\definecolor{SoftPeach}{rgb}{0.937,0.901,0.901}
\begin{table}[b]
\centering
\scriptsize
\caption{Simulation Configuration and Simulated Systems}
\vspace{-2mm}
\label{tab:simconfig}
\begin{tblr}{
  width = \linewidth,
  colspec = {Q[134]Q[620]},
  row{1} = {SoftPeach,c},
  row{15} = {SoftPeach,c},
  cell{1}{1} = {c=2}{0.94\linewidth},
  cell{3}{1} = {r=4}{},
  cell{7}{1} = {r=2}{},
  cell{9}{1} = {r=2}{},
  cell{15}{1} = {c=2}{0.94\linewidth},
  hline{1-3,12,14-16} = {-}{},
  hline{4-11,13,17-20,21,22,23} = {-}{},
  vlines,
}
\textbf{Baseline Virtuoso+Sniper Configuration} & \\
\textbf{Core} & 4-way Out-of-Order x86 2.9 GHz core\\
\textbf{MMU} & L1 I-TLB: 128-entry, 8-way assoc, 1-cycle latency\\
 & L1 D-TLB (4 KB): 64-entry, 4-way assoc, 1-cycle latency; L1 D-TLB (2 MB): 32-entry, 4-way assoc, 1-cycle latency \\ 
 & L2 TLB: 2048-entry, 16-way assoc, 12-cycle latency\\
 & 3-Page Walk Caches: 32-entry, 4-way, 2-cycle latency\\
\textbf{L1 Cache} & {L1 I/D-Cache: 32 KB, 8-way assoc, 4-cycle access latency}\\
 & LRU replacement policy;~IP-stride prefetcher~\cite{stride}\\
\textbf{L2 Cache} & 2 MB, 16-way assoc, 16-cycle latency\\
 & SRRIP replacement policy~\cite{srrip}; Stream prefetcher~\cite{streamer}\\
\textbf{L3 Cache} & 2 MB/core, 16-way assoc, 35-cycle latency\\
\textbf{DRAM} & 256 GB, DDR4-2400, $t_{RCD}$, $t_{CL}$=12.5 ns, $t_{RP}$=2.5 ns \\
\hline
\textbf{MimicOS} & Linux-like THP with 4 KB and 2 MB pages; HugeTLBFS;
                   Swap: 4 GB; Swapping threshold: 90\%;
                   Baseline fragmentation: 80\% \\
\hline
\textbf{Real System \newline(Validation)} & {Linux 5.15.0-60~\cite{linux-515}; DDR4-2400 Memory: 256 GB; \newline CPU: Intel Xeon Gold 6226R
2.90 GHz ~\cite{kratos20}}\\
\hline
\textbf{Simulated Systems Evaluated in Use Cases (\S\ref{sec:usecase1}-\ref{sec:usecase5})} &  \\
\textbf{Radix\newline~\cite{5-levelpaging-intel,5-levelpaging}} & {4-level tree; 4 KB page table frames; 3-Page Walk Caches (Physical Indexing): 32-entry, 2-way, 2-cycle} \\
\textbf{ECH~\cite{elastic-cuckoo-asplos20}} & {8K-entries/way; 4-way; Hash function: CITY~\cite{cityhash} 2-cycle}
 Perfect Cuckoo Walk caches for inter-page walks: 2-cycle   \\
\textbf{HDC~\cite{hash_dont_cache}} & Size: 4 GB; Open addressing; 8 PTEs/entry \\
\textbf{HT~\cite{powerpc-manual}} & Size: 4 GB; Chain Table; 8 PTEs/entry \\
\textbf{Utopia~\cite{kanellopoulos2023utopia}} & 2 x 8 GB RestSegs: 1$\times$4 KB pages and 1$\times$2 MB pages; RestSegs: 16-way, SRRIP replacement policy~\cite{srrip}; 1x FlexSeg with 4-level radix PT;
 TAR Cache: 8 KB, 2-cycle; SF Cache: 8 KB, 2-cycle \\
\textbf{Midgard~\cite{midgard}} & 64-entry L1 VLB: 1-cycle latency; 16-entry L2 Range-based VMA Lookaside Buffer: 4-cycle latency; B+ Tree for VMAs; 2-level MLB hierarchy; 6-level radix tree for M->P translation \\
\textbf{RMM~\cite{karakostas2015}} & 64-entry RLB: 9-cycle, Access in parallel with L2 TLB; Eager paging allocator with max order of 21; B+ Tree to store ranges \\
\end{tblr}
\end{table}

\noindent\textbf{Workloads.}
Table~\ref{tab:workloads} shows the benchmarks we used to evaluate Virtuoso. 
We select short-running applications ($<1$s) from various domains including Function-as-a-Service workloads~\cite{igniteMICRO23,Ustiugov_2021}, \konrevid{Large Language Model (LLM) inference}~\cite{llama,murty2024bagelbootstrappingagentsguiding,jiang2023mistral7b} and image processing~\cite{hamadard}.
We select long-running applications with high L2 TLB MPKI ($>5$) from the GraphBIG~\cite{Lifeng2015}, HPCC~\cite{luszczek_hpcc2006} and XSBench~\cite{Tramm2014} benchmark suites  which are also used by multiple prior works \konrevic{(e.g., ~\cite{elastic-cuckoo-asplos20,compendiaISMM2021,midgard,flataAsplos2022,kanellopoulos2023utopia,kanellopoulosMICRO2023victima}).}

\begin{table}[b]
  \centering
  \tiny
  \caption{Evaluated Workloads}
  \vspace{-2mm}
    \begin{tabular}{m{10em}m{21em}r}
    \toprule
    \textbf{\konrevic{Suite/Domain}} & \textbf{Workload} & \textbf{Data Set} \\
    \midrule
    GraphBIG~\cite{Lifeng2015} & Betweenness Centrality (BC), Breadth-first search (BFS) , Connected components (CC),
     Coloring (GC), PageRank (PR) , Triangle counting (TC) , Shortest-path (SP), k-Core (KC) & 50-100GB \\
    \midrule
    HPC & XSBench~\cite{Tramm2014}, randacc from  GUPS~\cite{Plimpton2006}  & 10 GB \\
    \midrule
    \konrevic{Function-as-a-Service} & AES, Image Resizing (IMG-RES), Word count of a document (WCNT), Database filter query (DB), JSON deserialization (JS) & <50MB \\
    \midrule
    \konrevic{Large Language Models} & Short-input short-output prompts using Llama 7B~\cite{touvron2023llamaopenefficientfoundation}, Bagel~\cite{murty2024bagelbootstrappingagentsguiding} and Mistral~\cite{jiang2023mistral7b} on top of llama.cpp~\cite{llama} & <2GB \\
    \midrule 
    Image Processing & 3D Hadamard Product~\cite{hamadard}, 3D Matrix Transposition~\cite{seismicNature}, 2D Matrix Sum & <2GB \\
    \bottomrule
    \end{tabular}
  \label{tab:workloads}%
\end{table}%

\subsection{Validation of Virtuoso}
\label{sec:validation}

\begin{figure}[h!]
    \centering
    \includegraphics[width=1.0\linewidth]{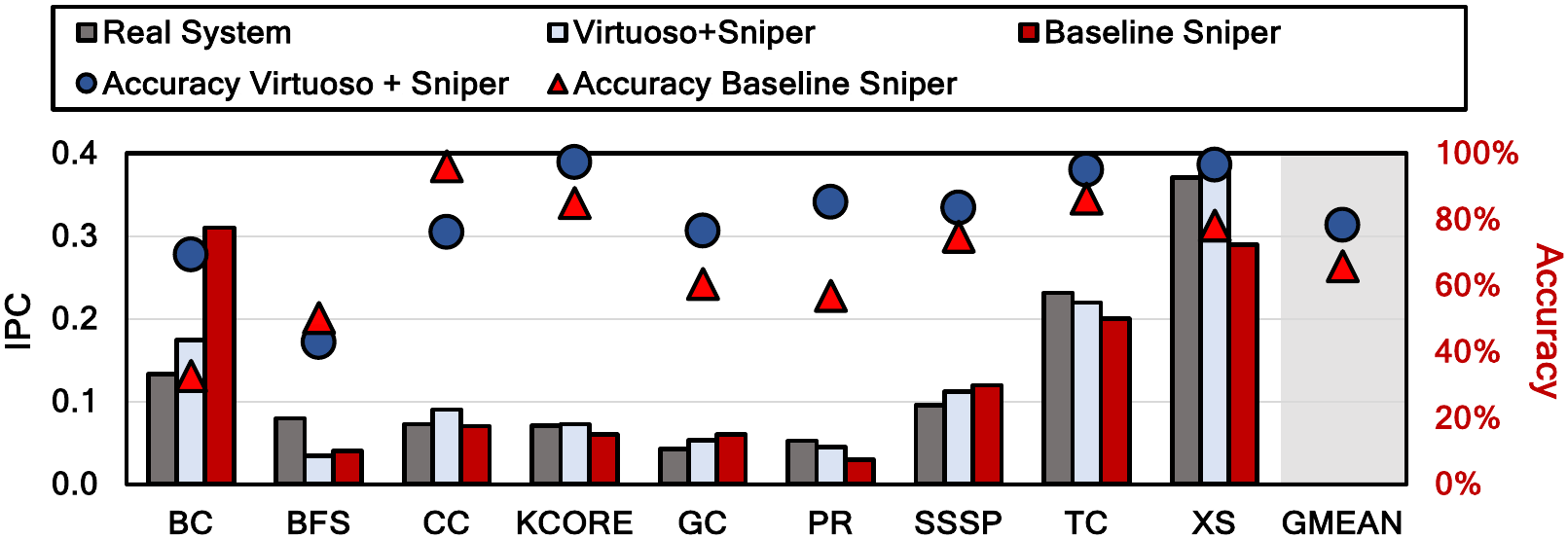}
    \vspace{-5mm}
    \caption{IPC \konrevid{estimation} accuracy estimation \konrevid{of} Virtuoso+Sniper and \konrevid{baseline} Sniper compared to a real system.}
    \label{fig:fault_virtuos}
\end{figure}

\head{IPC Validation} 
Figure~\ref{fig:fault_virtuos} shows the IPC performance estimation accuracy of Virtuoso+Sniper and baseline Sniper compared to a real system (\konrevid{Table~\ref{tab:simconfig}}) across
the long-running memory intensive workloads that are heavily affected by address translation.
Virtuoso (baseline Sniper) achieves 80\% (66\%) average accuracy in IPC estimation compared to the real system.
Virtuoso adapts to the dynamic characteristics of different workloads and achieves \konrevid{21}\% higher accuracy in IPC estimation \konrevid{versus} baseline Sniper which uses a fixed PTW latency (set as the average PTW latency obtained from a real system) regardless of the workload characteristics.

\begin{figure}[t!]
    \centering
    \includegraphics[width=1.0\linewidth]{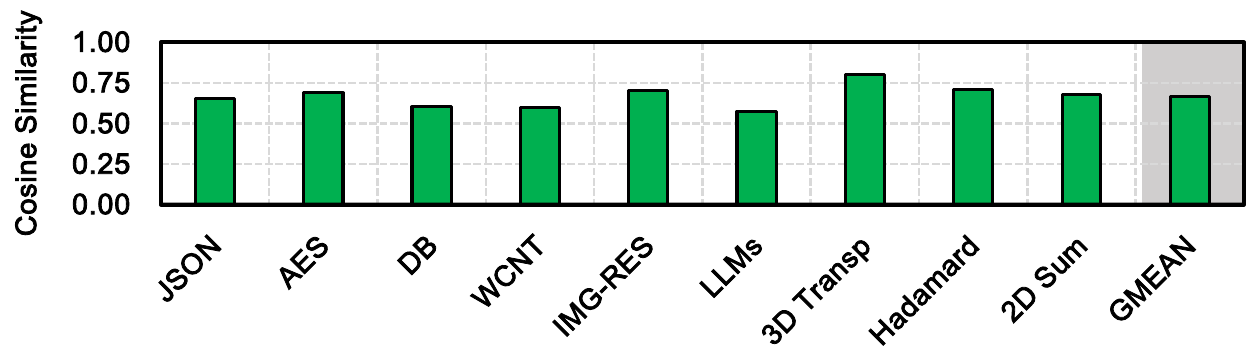}
    \vspace{-6mm}
    \caption{Cosine similarity between the page fault latency values measured by Virtuoso and the real system.}
    \label{fig:valid_pf}
\end{figure}

\head{Validation of Page Fault (PF) Latency}
We compare the PF latency reported by Virtuoso+Sniper against the page fault latency measured \konrevid{on} the real system.
We measure the real system PF latency at a fine granularity using \texttt{ftrace} and the \texttt{handle\_mm\_fault()} function tracer~\cite{ftrace}. Figure \ref{fig:valid_pf} shows the cosine similarity~\cite{cossim} of the PF latency reported by Virtuoso and the real system.\footnote{
We use the cosine similarity instead of the mean absolute error to account for the variance and the fluctuations in the PF latency across time.
} We use the short-running, page fault-bound workloads for which PF latency estimation is critical.
Despite using MimicOS, Virtuoso's userspace kernel that imitates only a subset of Linux kernel's memory management routines (\S\ref{sec:MimicOS}),
the cosine similarity of PF latency ranges from 60\% to 79\%, with an average of 66\% across all workloads.
We conclude that Virtuoso can approximate the PF latency with reasonable accuracy, even without modeling the entire Linux kernel.

\begin{figure}[h!]
    \centering
    \includegraphics[width=1.0\linewidth]{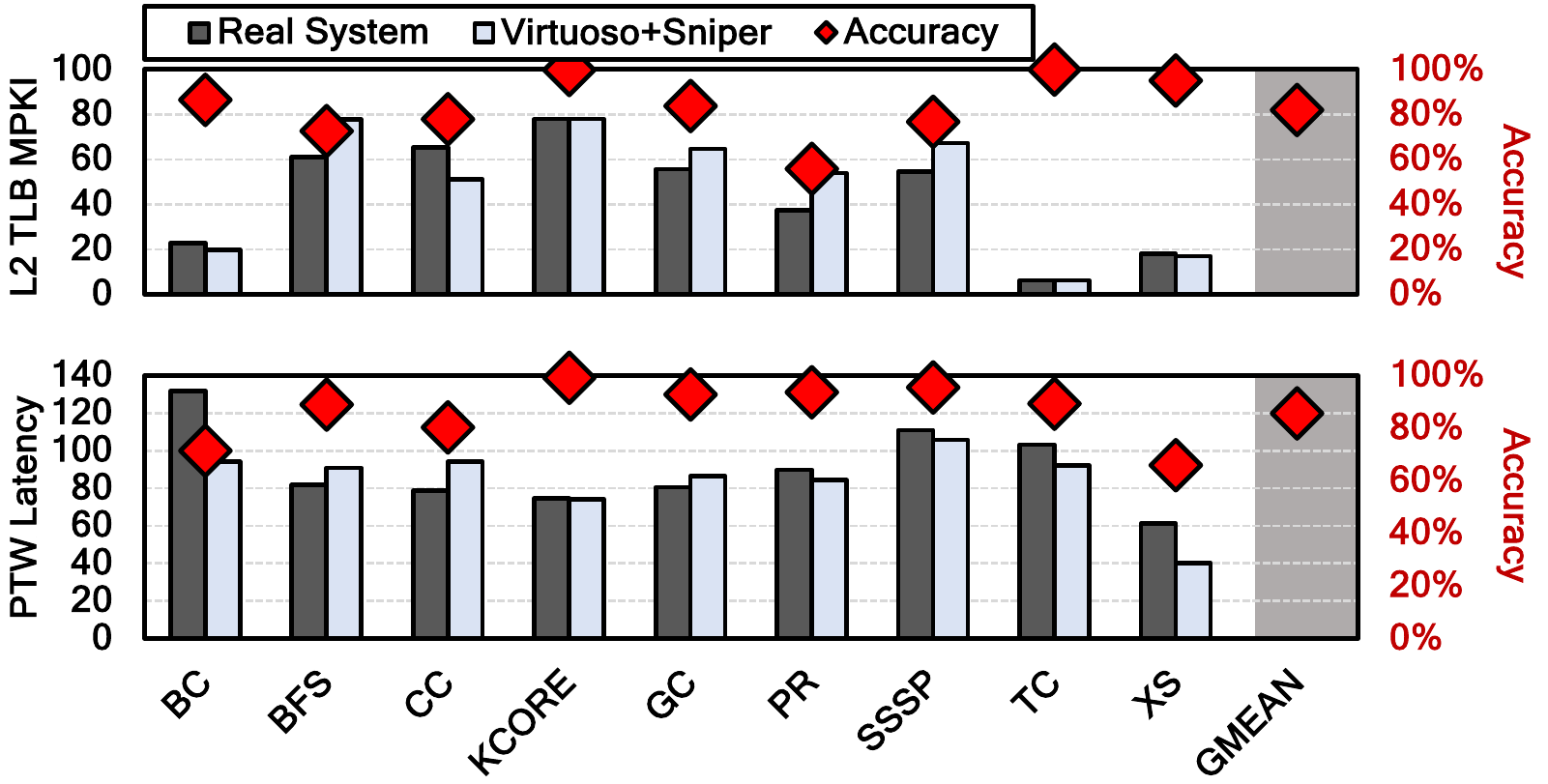}
    \vspace{-4mm}
    \caption{\konreve{(Top) L2 TLB MPKI and (Bottom) PTW latency reported by Virtuoso+Sniper compared to a real system.}}
    \label{fig:valid_ptw}
\end{figure}

\head{Validation of MMU Performance}
Figure~\ref{fig:valid_ptw} shows the L2 TLB misses per kilo instructions (MPKI) and the PTW latency of Virtuoso+Sniper compared to the real system.
For this experiment, we use the long-running workloads that are heavily affected by address translation \konrevid{latency} and \konrevid{thus} by the effectiveness of the MMU. 
We observe that Virtuoso estimates the L2 TLB MPKI and the PTW latency with, on average, 82\% and 85\% accuracy, respectively. 
Virtuoso accurately models the MMU performance of the real system, 
which is essential for capturing the address translation overheads in data-intensive workloads. 

\subsection{Simulation Time and Memory Overhead }
\label{sec:simcomp}

\begin{figure}[h!]
    \centering
    \includegraphics[width=1.0\linewidth]{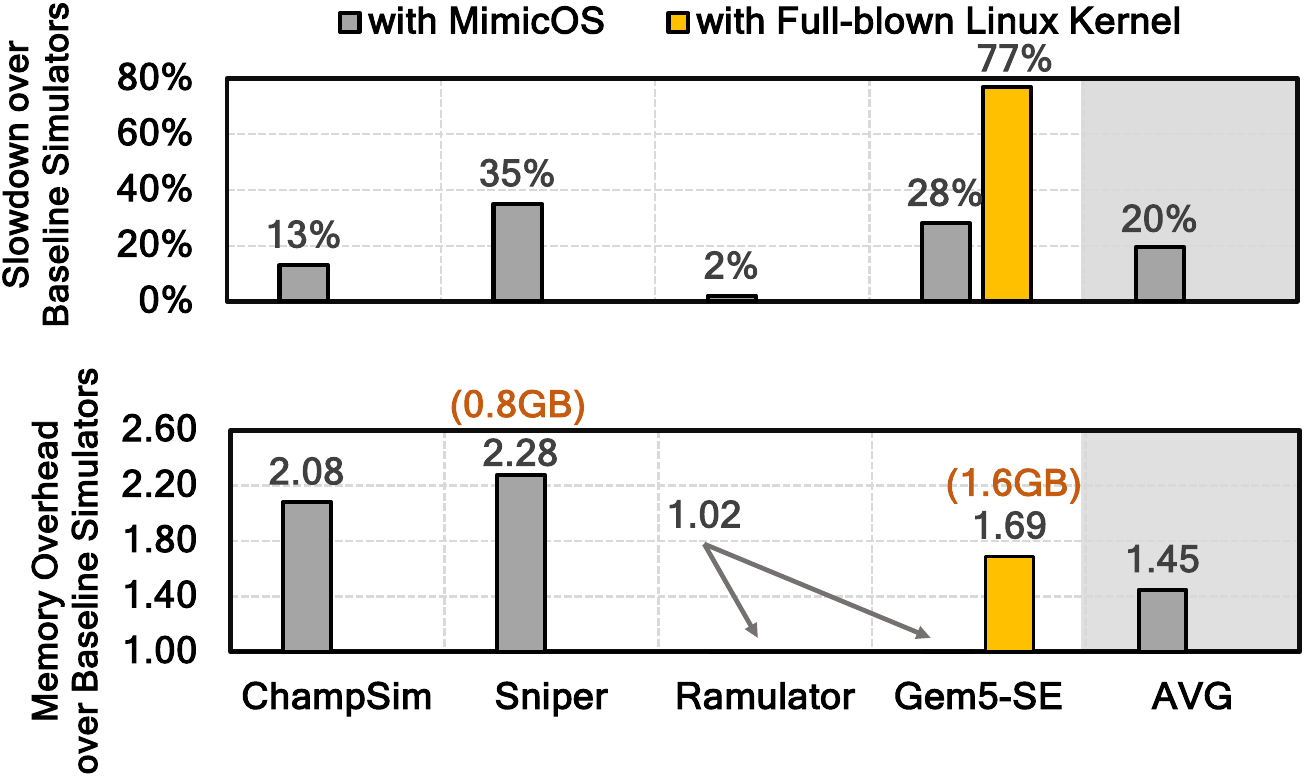}
    \vspace{-5mm}
    \caption{\konreve{Simulation time and memory usage overheads of \konrevid{integrating MimicOS into Sniper}, ChampSim, Ramulator and gem5-SE compared to their baseline versions and gem5-FS.}}             
    \label{fig:virtuoso_sims}
\end{figure}

Fig.~\ref{fig:virtuoso_sims} shows the simulation time and memory consumption overhead when we \konrevif{integrate} MimicOS \konrevid{into} Sniper, ChampSim, Ramulator, and gem5-SE compared to their baseline versions and gem5-FS.
We report worst-case overheads using \texttt{randacc}, which incurs the highest \konrevid{number} of page faults per kilo instructions \konrevid{(PFKI)} and ultimately frequent MimicOS-simulator communication.
We make five key observations. First, \konrevid{integrating} MimicOS increases simulation time by an average of 20\% due to additional simulated instructions. 
Second, enabling full-system mode in gem5 leads to a 77\% increase in simulation time \konrevid{compared to gem5's syscall-emulation mode.}
Third, using MimicOS results in a 1.45x average increase in memory consumption across all simulators.
\konrevid{Fourth, in ChampSim and Sniper, we observe nearly ~2.1x memory overhead since we enable online binary instrumentation for MimicOS.
On the contrary, in Ramulator where we use offline binary instrumentation and in gem5 where we reuse the existing binary emulation infrastructure, MimicOS leads to only ~1.02x overhead.}
Last, in terms of raw memory usage, porting MimicOS to Sniper leads to 0.8GB memory usage,  
whereas gem5-FS consumes double (1.6GB), leading to up to 2x lower simulation job throughput when memory capacity is limited.

\begin{figure}[h!]
    \centering
    \includegraphics[width=1.0\linewidth]{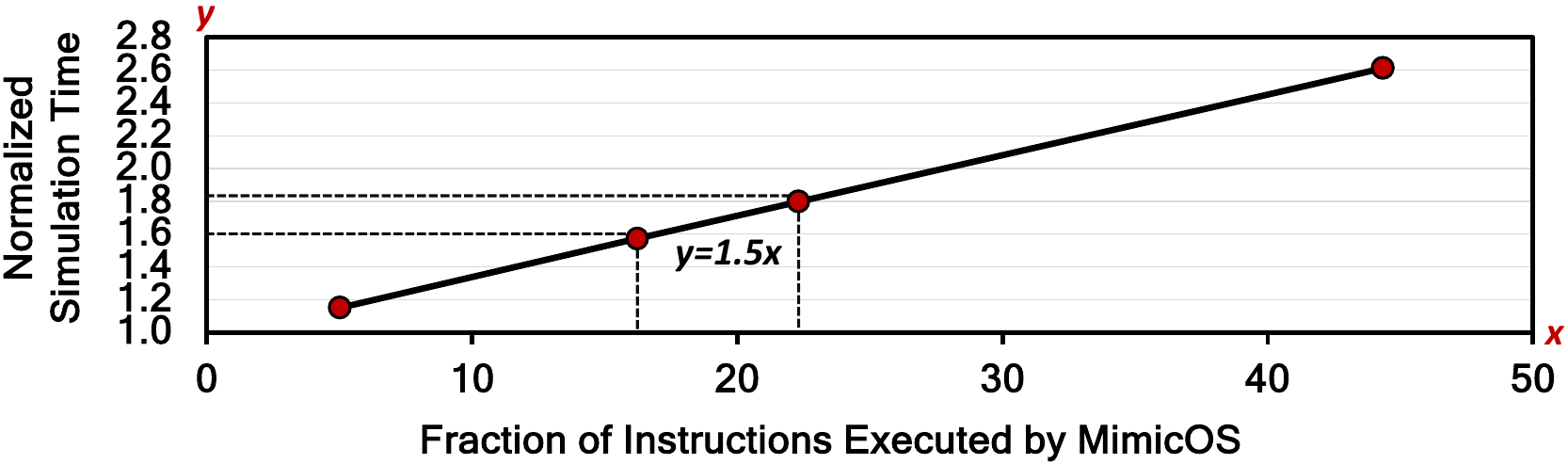}
    \caption{\konreve{Correlation between the number of instructions executed by MimicOS and the simulation time overhead.}}
    \label{fig:virtuoso_sim_corr}
    \vspace{-1mm}
\end{figure}

\textbf{Correlation Between Simulation Time and Number of MimicOS Instructions.}
Figure~\ref{fig:virtuoso_sim_corr} shows the correlation between the number of MimicOS instructions and the simulation time overhead when \konrevid{we integrate} MimicOS with Sniper.  
To perform this analysis, we crafted a microbenchmark where the number of MimicOS instructions is varied while keeping the total number of simulated instructions constant.
We observe a strong correlation between the number of MimicOS instructions and the simulation time overhead 
across all simulation points. As the number of MimicOS instructions increases, the simulation time overhead also increases, by a factor of 1.5x on average.
We also verify this trend for gem5-SE and gem5-FS \konrevid{(see extended version~\cite{virtuoso_arxiv})}.

\subsection{\nb{Use Case 1:} Alternative Page Table Designs}
\label{sec:usecase1}

\nb{We} evaluate different page table (PT) designs to draw insights on the trade-offs between address translation latency, memory interference and page fault latency.
We evaluate the following designs:
(i) \textbf{Radix}: a 4-level radix-based PT design~\cite{intelx86manual} and Linux-like THP enabled~\cite{corbet2011,corbet2017},
(ii) \textbf{ECH}: elastic cuckoo hash PT design~\cite{elastic-cuckoo-asplos20},
(iii) \textbf{HDC}: 4GB global open-addressing-based hash PT ~\cite{hash_dont_cache}, \nb{and}
(iv) \textbf{HT}: 4GB global chain-based hash PT ~\cite{powerpc-manual}. 
In this use case, we define memory fragmentation as the percentage of free 2MB pages compared to the total number of 2MB pages.

\begin{figure}[h!]
    \centering
    \includegraphics[width=1.0\linewidth]{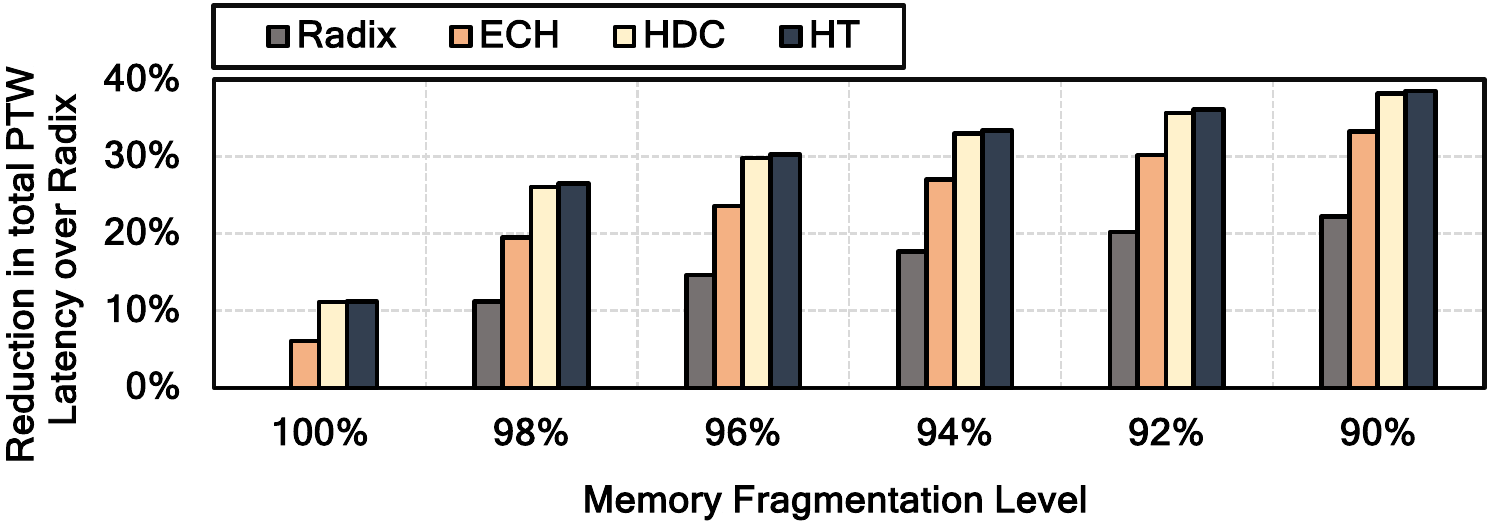}
    \caption{\konrevid{Reduction in \konrevif{total} PTW latency achieved by hash-based \konrevif{PTs} compared to Radix across different memory fragmentation levels.}}
    \label{fig:ptw_frag}
\end{figure}

\begin{figure}[t!]
    \centering
    \includegraphics[width=1.0\linewidth]{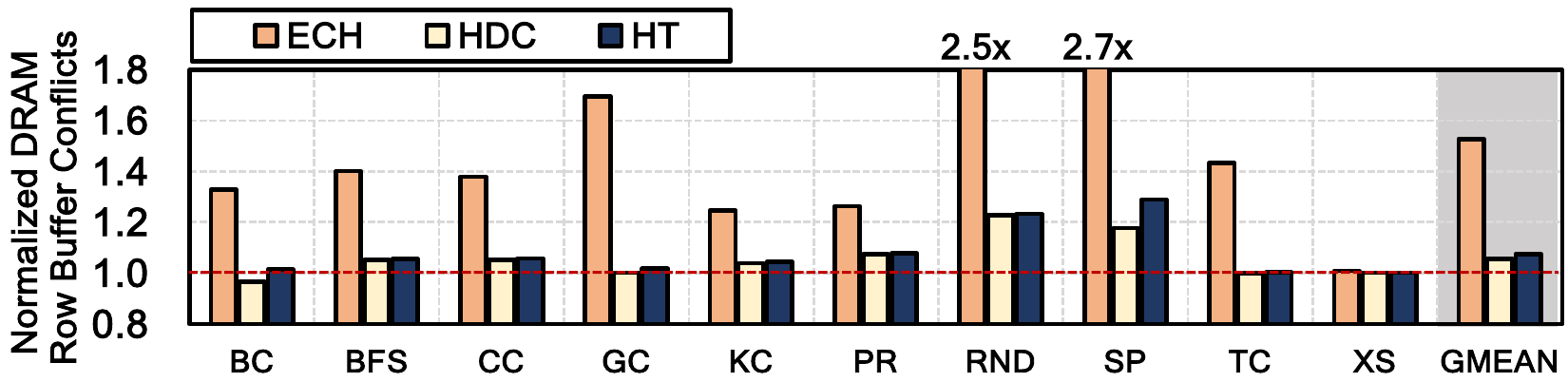}
    \caption{\konrevid{Normalized DRAM row buffer conflicts for ECH, HDC and HT over Radix.}}
    \label{fig:ptsconflicts}
\end{figure}

\textbf{Effect of PT Design on Translation Latency \& Memory Interference.}
Figure~\ref{fig:ptw_frag} shows the \konrevid{reduction in \konrevif{total} PTW \nb{latency} achieved by ECH, HDC and HT compared to Radix, across different memory fragmentation levels.}
We make two key observations. First, all three hash-based PT designs 
consistently reduce \konrevif{the total} PTW latency over \konrevid{Radix} across all \konrevid{memory} fragmentation levels.
Second, the reduction in \konrevif{total} PTW latency achieved by all hash-based PT designs increases with decreasing fragmentation levels.
To better understand the effect of PT design on the system, in Figure~\ref{fig:ptsconflicts} we show the total DRAM row buffer conflicts (induced by activating rows that contain either data or page table entries) of ECH, HDC, and HT compared to Radix.
We observe that ECH increases total DRAM row buffer conflicts by $52\%$ \konrevid{over Radix} while HDC and HT reduce DRAM row-buffer conflicts by $5\%$ and $7\%$, respectively. 
\konrevid{Probing ECH during a PTW requires multiple memory accesses (one access for each Cuckoo nest in the hash table), causing high interference in the main memory.}

\textbf{Effect of PT \konrevif{Design} on Minor Page Fault Latency (MPF).} PT design can significantly impact MPF latency due to differences in \konrevid{PT update or insertion operations}. 
For example, Radix requires up to 4 memory accesses to insert a new entry, while ECH may require 1 or more depending on load or insertion order. 
Figure 15 shows \konrevid{the reduction in \konrevif{total} MPF latency achieved by the hash-based PTs over Radix. We make two key observations. First, ECH, HDC, and HT respectively reduce MPF latency 
 by 9\%, 18\% and 19\%, on average across all workloads.} This occurs because hash-based PTs are allocated (or expanded) with large physical memory chunks compared to Radix that allocates 4KB frames on-demand.
Second, HDC and HT reduce MPF latency across all workloads, while ECH \emph{increases} it in RND due to multiple memory accesses \konrevid{caused by} \ak{hash} collisions.

\observation{\sepherd{Although \konrevie{ECH} reduces the latency of \konrevie{PTWs}, it causes higher main memory contention and \konrevid{sometimes} increases the latency of MPFs compared to a radix-based baseline.}}

\begin{figure}[h]
    \centering
    \includegraphics[width=1.0\linewidth]{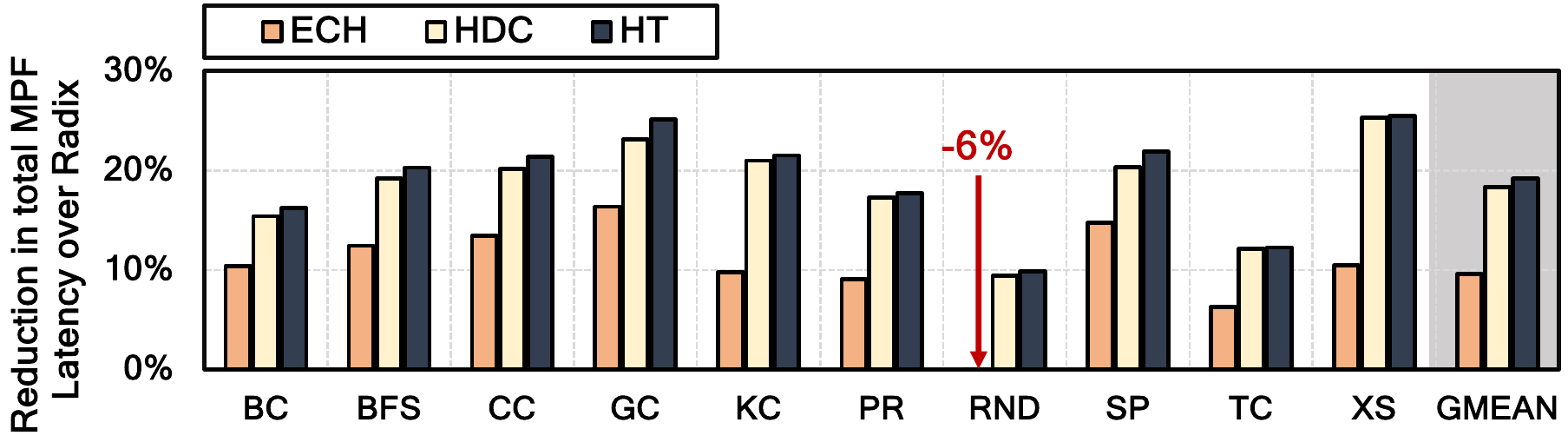}
    \caption{\konrevid{Reduction in \konrevif{total} minor page fault (MPF) latency achieved by hash-based PTs compared to Radix.}}
    \label{fig:pt_fault_lat}
\end{figure}

\subsection{\nb{Use Case 2:} Physical Memory Allocation in LLMs}
\label{sec:usecase2}

\nb{We} examine \nb{the effect of} different physical memory allocation policies:
(i) \textbf{BD}: a buddy allocator that only provides 4KB pages and updates the PT accordingly,
(ii) \textbf{CR-THP}: a conservative reservation-based THP allocator~\cite{reserve} that reserves a 2MB physical memory region upon the initial allocation of a 4KB page, and fully upgrades it to a 2MB page once over 50\% of the 4KB pages within that region are allocated,
(iii) \textbf{AR-THP}: an aggressive reservation-based THP allocator~\cite{reserve} that reserves a 2MB physical memory region upon the initial allocation of a 4KB page, and fully upgrades it to a 2MB page once over 10\% of the 4KB pages within that region are allocated, and
(iv) \textbf{UT}: a \konrevid{Utopia~\cite{kanellopoulos2023utopia}} system \nb{with} memory segments of different sizes (4MB, 32MB, 512MB) and associativity (8,16) that \konrevid{employ} a restrictive hash-based virtual-to-physical address mapping.

\begin{figure}[h]
    \centering
    \includegraphics[width=1.0\linewidth]{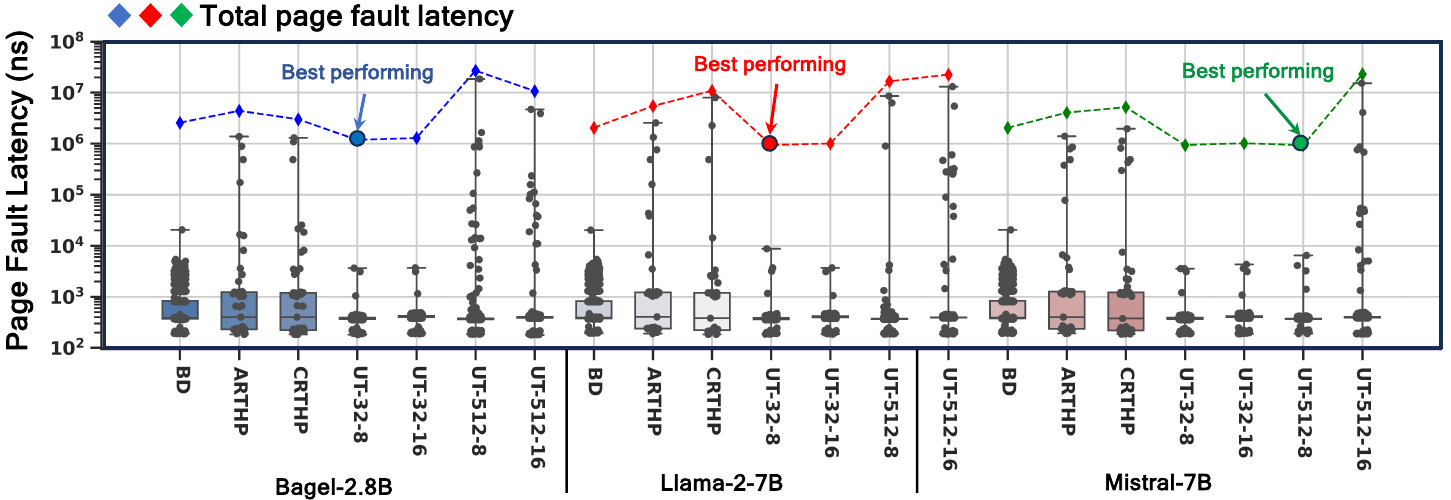}
    \caption{Page fault latency distribution \konrevif{with seven different physical memory allocation policies for} three LLM workloads.}
    \label{fig:allocators}
\end{figure}

Figure~\ref{fig:allocators} shows the PF latency distribution across all allocation policies in three LLM inference workloads. We make three observations.
First, THP-based allocators (CR-THP and AR-THP) show similar median latency to BD but with a >1000x increase in tail latency.
Second, UT-32MB/16-way achieves the lowest PF latency as it provides large contiguous segments for \konrevid{fast} hash-based page allocations. 
Third, as we increase the restrictive segment size (e.g., UT-512MB/16-way) both the total and tail PF \konrevid{latencies} increase compared to UT-32MB/16-way. 
This is because, allocating data in a very large segment limits the spatial locality of the \konrevid{data structure that stores the allocation metadata} (i.e., virtual tags for each physical page)
which in turn increases PF latency. %

\observation{\sepherd{Restricting the virtual-to-physical address mapping leads to faster page fault handling due to the lightweight hash-based page allocation routine.}}

\subsection{Evaluating Different MMU Designs}

\nb{We} draw insights into how different MMUs affect {microarchitectural and system-level metrics.}
We evaluate the following designs:
(i) \textbf{Utopia}~\cite{kanellopoulos2023utopia}: a system equipped with a 16GB-large physical memory segment that \konrevid{employs} a restrictive address mapping,
(ii) \textbf{RMM}~\cite{karakostas2015}: a system that employs, on the software side, eager paging to allocate large contiguous physical segments and, on the hardware side, a range lookaside buffer and range walker to quickly retrieve contiguity information,
(iii) \textbf{Midgard}~\cite{midgard}: a system that employs an intermediate address space and two-level address translation, with a frontend that employs two VMA lookaside buffers 
and a backend that employs a 4-level radix tree.
\konrevb{\konrevid{We} define memory fragmentation based on the underlying design: for Utopia, we define memory fragmentation as the number of available 2MB pages, 
including the contiguous 2MB pages needed to form the RestSeg, compared to the total number of 2MB pages.
For RMM, we define memory fragmentation as the ratio of \konrevid{the total size of} the top 50  largest \konrevie{unallocated} contiguous segments  to the total main memory size.
For Midgard, we define memory fragmentation as the
\konrevid{number} of 2MB pages that are available for allocation for the backend translation level compared to the total number of 2MB pages.}

\subsubsection{\nb{Use Case 3:} Intermediate Address Space Schemes} 
\label{sec:usecase3}

\begin{figure}[h!]
    \centering
    \includegraphics[width=1.0\linewidth]{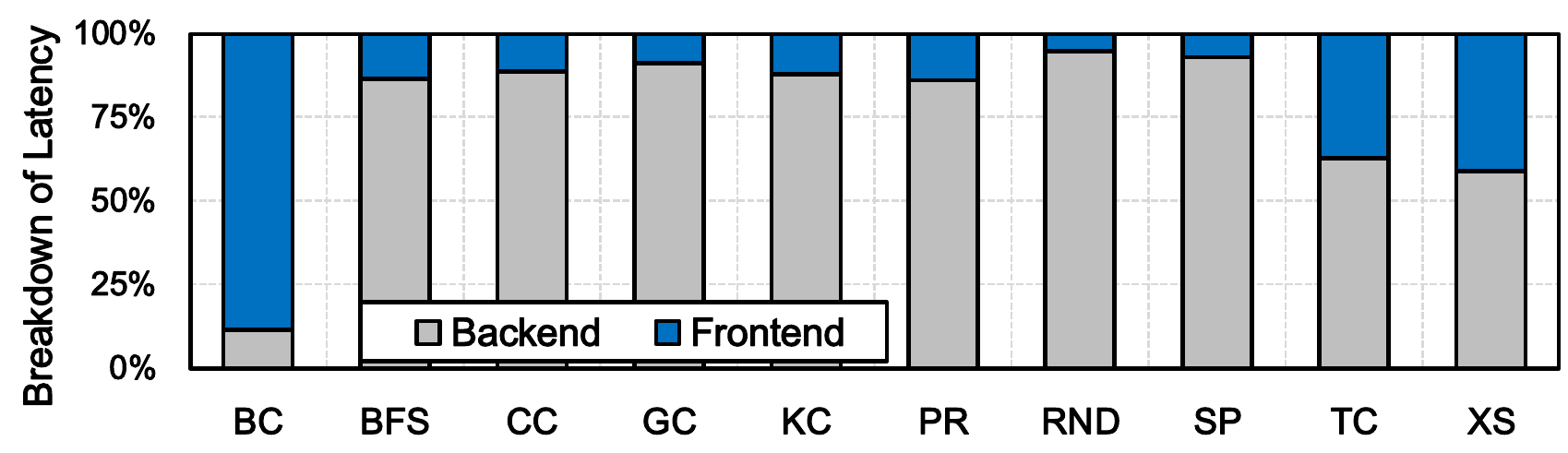}
    \caption{Breakdown of \konrevie{translation} latency in Midgard.}
    \label{fig:midgard}
\end{figure}

Figure~\ref{fig:midgard} shows the breakdown of address translation latency in Midgard~\cite{midgard} to understand the effects of frontend and backend address translation.
We make two key observations. First, most workloads spend less than 20\% of the total translation latency in the frontend translation since they use 
a small number of large VMAs. Hence, the frontend lookaside buffers can effectively cache all the VMA information. 
Second, we observe that \texttt{BC} spends more than 50\% of the total translation latency in the frontend.

\begin{figure}[h!]
    \centering
    \includegraphics[width=1.0\linewidth]{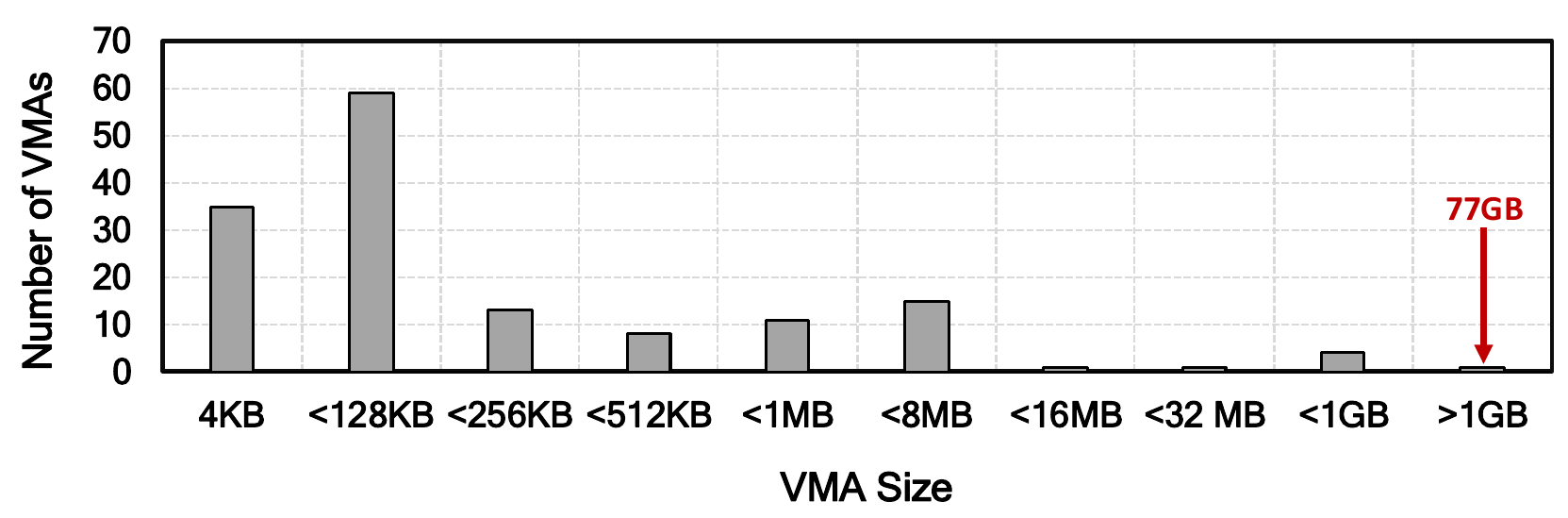}
    \vspace{-6mm}
    \caption{\konrevif{Number of VMAs of different sizes in BC.}}
    \label{fig:vmas_bc}
\end{figure}

To better understand this phenomenon, we investigate the number and size of virtual memory areas (VMA)~\cite{vma} involved in \texttt{BC}. As shown in Figure~\ref{fig:vmas_bc}, BC uses (i) one VMA occupying 77GB of VA space and (ii) 147 smaller VMAs ranging from 4KB to 1GB. 
While the large VMA is efficiently cached in the frontend VMA lookaside buffers, the 147 smaller VMAs are not covered efficiently by either the L1 or L2 VMA-LBs (3\% hit ratio in L2 VLB), resulting in high frontend translation latency. 
We conclude that Midgard's frontend design needs further optimization to handle workloads with many small VMAs, despite the large VMAs being efficiently cached.

\observation{\sepherd{Schemes that employ intermediate address spaces can be further optimized to reduce the frontend translation latency for workloads with a large number of small VMAs.}}

\subsubsection{\nb{Use Case 4:} Restricting the VA-to-PA Mapping}
\label{sec:usecase4}

\begin{figure}[h!]
    \centering
    \includegraphics[width=1.0\linewidth]{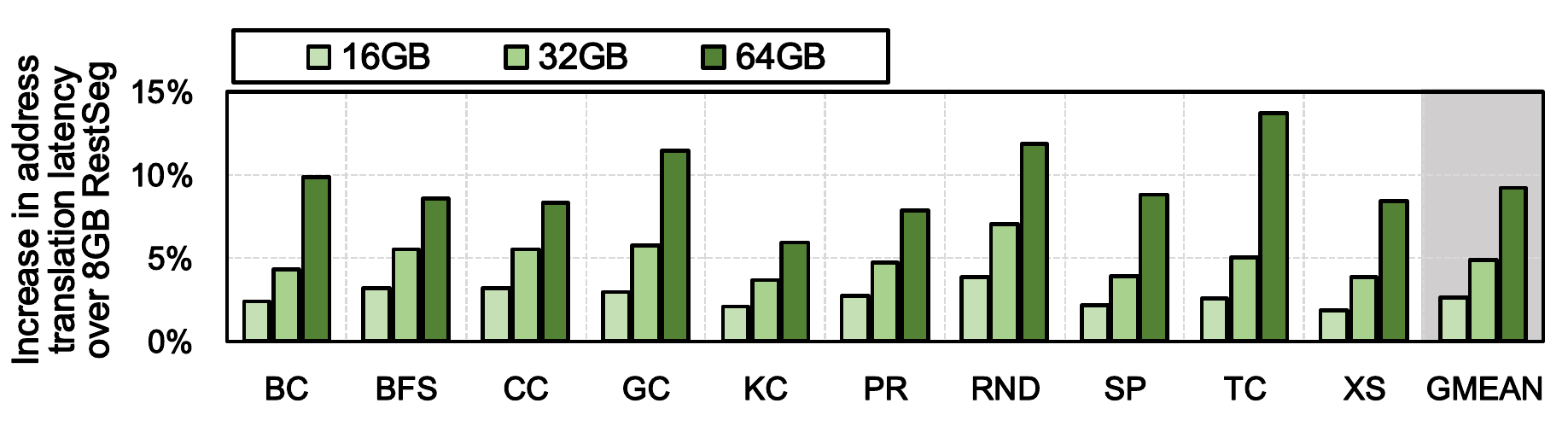}
    \vspace{-5mm}
    \caption{Increase in translation latency achieved by increasing the RestSeg size over Utopia with an 8GB RestSeg.}
    \label{fig:utopia_size}
\end{figure}

We evaluate the effects \ak{of the size of the restrictive segment} (RestSeg) in Utopia~\cite{kanellopoulos2023utopia}. 
Figure~\ref{fig:utopia_size} shows the increase in translation latency as we increase the Utopia RestSeg size up to 64GB compared to Utopia that employs an 8GB RestSeg.
We draw the following insight: as we increase the size of the RestSeg, address translation latency increases, up to 10\% for the largest RestSeg compared to the 8GB RestSeg. 
This is because a large RestSeg increases the latency of accessing address translation metadata (RSW as described in~\cite{kanellopoulos2023utopia}).

\observation{\sepherd{Selecting the size of a memory segment that enforces a restrictive \konreve{VA-to-PA} mapping poses a trade-off: larger segments reduce the frequency of page table walks for data within these segments, yet they may increase address translation latency.}}

\begin{figure}[h!]
    \centering
    \includegraphics[width=1.0\linewidth]{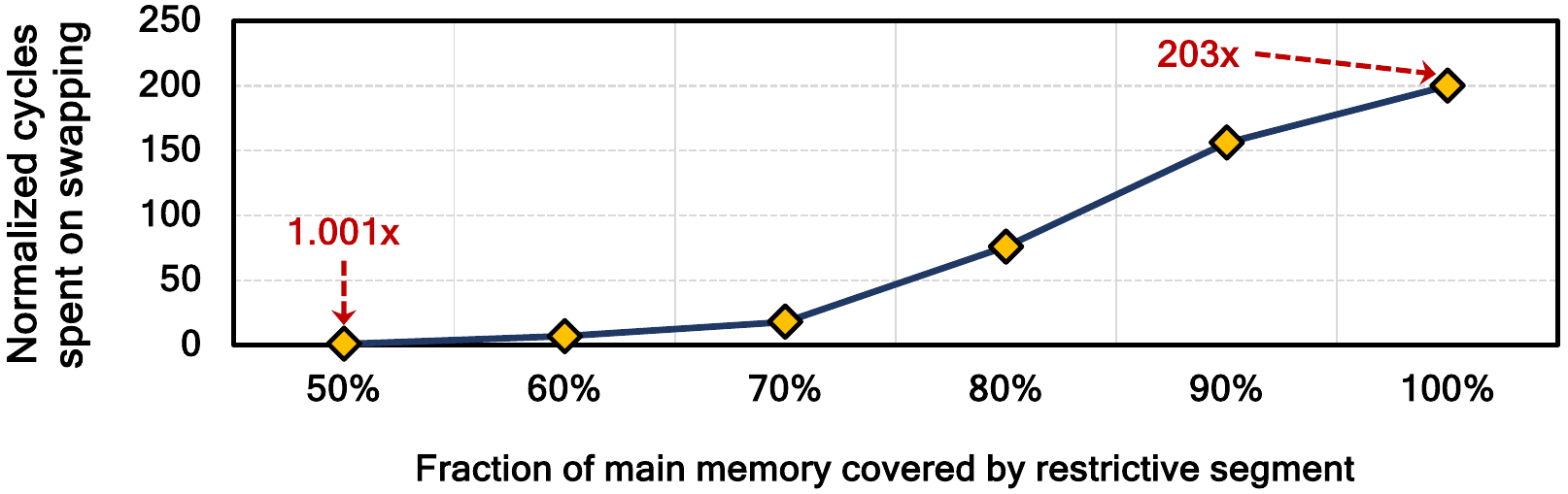}
    \caption{\konrevid{Time spent in swapping activity for different restrictive segment sizes (in Utopia),
    normalized to Radix.}}
    \label{fig:swap_utopia}
\end{figure}

\head{Effect of Utopia on \konrevid{Swapping Activity}}
We evaluate the effect of Utopia on swapping activity using a setup where Virtuoso \konrevid{is integrated into} Sniper~\cite{sniper} \emph{and} MQSim~\cite{tavakkol2018mqsim}. In this setup, Utopia is configured with restrictive segments capturing large portions of main memory (>50\%), and we measure the time spent swapping in/out of memory. When memory usage exceeds 90\%, the system begins swapping pages to disk. Figure \ref{fig:swap_utopia} shows the normalized time spent in swapping for different restrictive segment sizes compared Radix.
We observe that swapping time increases with larger restrictive segments, \konrevid{reaching up to 203x for the largest size
compared to Radix}. This occurs because restrictive segments cause hash collisions that prevent data from being stored in memory even in the presence of free space. 
Thus, careful selection of restrictive segment size is crucial to minimize swapping overheads.

\observation{\sepherd{Enforcing a restrictive hash-based mapping across very large memory segments leads to increased swapping activity}.}

\subsubsection{\nb{Use Case 5:} Exploiting Contiguity Information}
\label{sec:usecase5}

\begin{figure}[h!]
    \centering
    \includegraphics[width=1.0\linewidth]{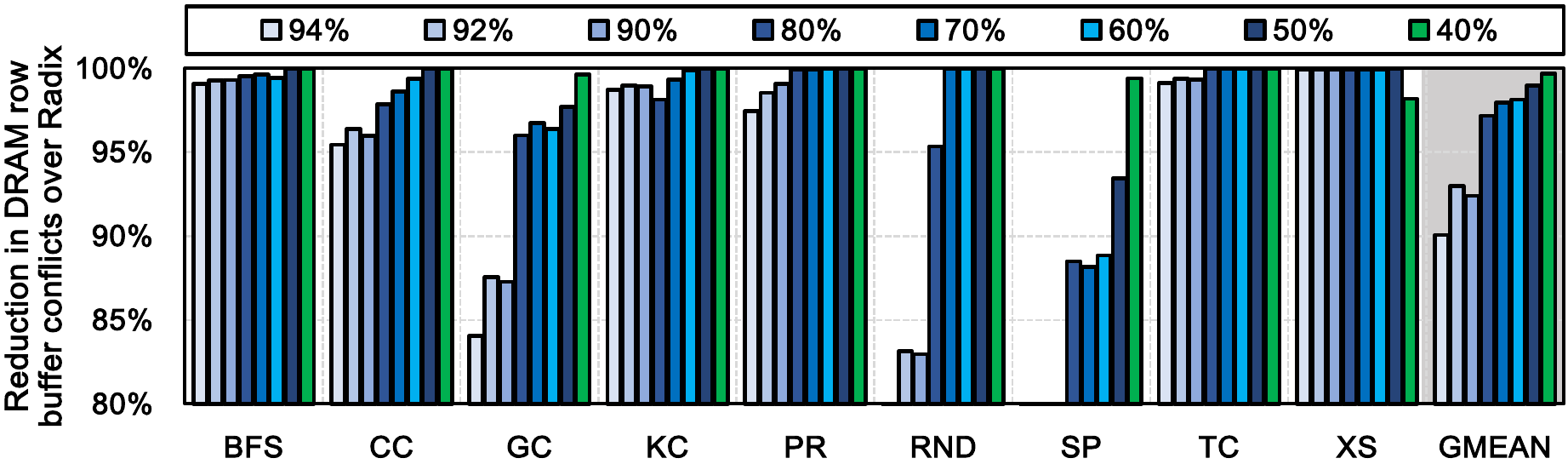}
    \caption{\konrevid{Reduction in DRAM row buffer conflicts (caused by address translation metadata) achieved by RMM, over Radix, across different \konrevif{memory} fragmentation levels.}}
    \label{fig:dramrb_rmm}
\end{figure}

We further explore the effect of memory fragmentation on \konrevid{exploiting virtual-to-physical address} contiguity  
to reduce PTWs as described in RMM~\cite{rmmisca15}. Figure~\ref{fig:dramrb_rmm} shows the \konrevid{reduction in} DRAM row buffer conflicts caused by address translation metadata (contiguity information and page table entries) \konrevid{achieved by} RMM over Radix, across different fragmentation levels  
We observe that even with 94\% fragmentation, RMM reduces DRAM row buffer conflicts caused by address translation metadata by $90\%$ on average over Radix due to the reduced number of PTWs.

\observation{\sepherd{Even at \konrevid{mid-to-high} memory fragmentation levels, employing contiguity-based schemes significantly reduces DRAM row buffer conflicts caused by page table accesses.}}

\section{Related Work}

\konrevic{To our knowledge, Virtuoso is the first simulator that bridges the gap between emulation-based and
full-system simulators enabling accurate exploration of VM designs in a fast and flexible way.}
Various simulators (\konrevic{e.g.,}\Allsimulators) and simulation methodologies (\konrevic{e.g.,}~\cite{interval,vsim,sampledsimul,barrierpoint,timetravel,scalemodel,liu2023pacsim,elfies,looppoint,badgertrap}) have been developed to model different system components. 
In \S\ref{sec:motivation}, we examine the key characteristics of emulation-based and full-system simulators and compare them against Virtuoso.
In this section, we discuss other related simulation methodologies and provide a broad overview of works that focus on VM optimizations.

\subsection{\konrevic{First-Order Models}}

First-order models, combined with instrumentation tools (e.g., BadgerTrap~\cite{badgertrap}), \konrevic{are} used in prior VM research (\konrevic{e.g.,} ~\cite{rmmisca15,partialMICRO2020,hash_dont_cache})
to approximate VM overheads . These models are typically analytical (e.g., fixed latency for PTW) which makes them valuable for quickly estimating the performance impact of new VM features. However, they overlook critical \konrevic{dynamic} effects arising from hardware and OS interactions, such as the volume of page table data stored in caches, DRAM contention due to page table accesses, and large page availability affected by fragmentation. These effects exhibit dynamic behavior and can significantly influence evaluation results.

In contrast, Virtuoso captures both \konrevic{first-order and dynamic} effects in VM performance analysis. For instance, as demonstrated in \S\ref{sec:usecase1}, Virtuoso measures first-order metrics (e.g., page table walk latency, page fault latency) alongside \konrevic{dynamic} effects (e.g., resource contention) of page table design. Thus, Virtuoso serves as an alternative for simulating hardware/OS interactions at higher detail when necessary.

\subsection{FPGA-Accelerated Simulation}
Several prior works explore FPGA-based approaches to accelerate system simulation (\konrevic{e.g.,} \SimFPGA). FireSim~\cite{firesim} is an FPGA-accelerated platform that enables fast, cycle-exact simulation of large-scale systems, such as server blades.  FAST~\cite{fastMICRO2007} is a hybrid FPGA-CPU simulator that offloads its timing model computation on an FPGA while executing the functional model on a CPU.

FPGA-accelerated simulators come with notable challenges: (i) porting simulation models to Register-transfer level (RTL) requires substantial development effort and time, (ii) slow compilation due to RTL synthesis, and (iii) existing FPGA-based prototypes may not fully represent modern systems due to constraints such as discrepancies between FPGA and DRAM operating frequencies.
While these simulators provide \konrevic{fast and accurate simulation}, they can be impractical for rapid prototyping \konrevic{(and programming)} in fast-evolving \konrevie{HW/SW} environments, such as virtual memory solutions.
Compared to FPGA-accelerated simulators, Virtuoso prioritizes ease of development, use and versatility while providing relatively high simulation speed and \akdel{high} accuracy. %

\subsection{Simulating Large-Scale Memory/Storage Systems}

\konrevic{Prior works optimize how program values are stored by the simulator\konrevid{,} enabling large-scale memory and storage system simulation (e.g., ~\cite{david,zero_sim,exalt}).
David~\cite{david} and Exalt~\cite{exalt} employ semantics-aware data representation schemes that lead to highly-efficient data compression, enabling large-scale storage simulation.}
Øsim~\cite{zero_sim} models large-scale memory systems on commodity hardware by leveraging the observation that most data-intensive workloads follow similar control flows, enabling efficient memory compression. 
Virtuoso can be integrated with these simulators to model real program values while optimizing memory usage.

\subsection{Virtual Memory Optimizations}
\sepherd{
To improve VM, prior works explore several key approaches:
(i) enabling large page sizes (\konrevic{e.g.,} \VMlargepages),
(ii) enforcing virtual-to-physical address contiguity to increase the processor's address translation reach (\konrevic{e.g.,} \VMcontiguity),
(iii) employing restrictive virtual-to-physical address mappings (\konrevic{e.g.,} \VMrestrictive),
(iv) designing alternative page table structures to reduce PT walk latency (\konrevic{e.g.,} \VMpagetable),
(v) employing TLB prefetching (\konrevic{e.g.,} \VMtlbprefetching),
(vi) optimizing TLB replacement policies (\konrevic{e.g.,} \VMtlbreplacementpolicy),
(vii) storing TLB entries in the cache to minimize PT walks (\konrevic{e.g.,} \VMtlbincache),
(viii) leveraging hardware support to reduce page fault handling latency (\konrevic{e.g.,} ~\cite{hbdpISCA2020,minorfaultTACO2022,mementoMICRO2023}), 
(ix) employing hardware mechanisms to accelerate PT walks (\konrevic{e.g.,} \VMpwcs),
(x) optimizing VM components for efficient address translation in virtualized environments (\konrevic{e.g.,} \VMvirtualized) and
(xi) employing intermediate address spaces to defer address translation (\konrevic{e.g.,} \VMintermediate).
Developing these techniques requires extensive simulation effort at both the OS and the hardware model levels. Virtuoso provides a comprehensive toolset of state-of-the-art VM techniques, offering a common ground that makes it easier to develop and evaluate existing and new VM solutions.
}

\section{Conclusion}

\konrevic{We introduced Virtuoso, a \konrevic{new simulation methodology} that enables quick and accurate prototyping and evaluation of virtual memory (VM) schemes.}
Virtuoso's key idea is to employ a lightweight userspace kernel written in a high-level language, which comprises \konrevic{of} a subset of the OS's VM-related functionalities to: (i) accelerate simulation, (ii) \konrevic{simplify the development of new OS routines}, and (iii) \konrevic{accurately}  \konrevid{evaluate} different VM schemes. \konrevic{We integrate Virtuoso} with five architectural simulators and validate it against a real high-end server-grade CPU. \konrevia{To showcase Virtuoso's versatility, we conduct five case studies demonstrating its applicability to various VM research areas.} \konrevic{Our evaluation demonstrates that Virtuoso provides a new point in the design space of simulators that strikes a unique balance between \konrevid{simulation speed, accuracy\konrevif{,} and versatility.}
We conclude that Virtuoso can become a useful platform for researchers to implement, compare and evaluate new and existing VM designs. To enable further research, we make Virtuoso freely available at \url{https://github.com/CMU-SAFARI/Virtuoso}.}

\section*{Acknowledgements}

We thank the anonymous reviewers of MICRO 2024 and ASPLOS 2025 for their feedback and
the SAFARI Research Group members for
providing a stimulating intellectual environment.
\konrevif{We thank Ian Ganz for his help during early stages of this work.} We acknowledge
the generous gifts from our industrial partners: Google, Huawei,
Intel, Microsoft, and VMware, \konrevif{and 
the Semiconductor Research Corporation. This work was supported in part by the ETH Future Computing Laboratory.}

\bibliographystyle{unsrt}
\balance 
\bibliography{refs,daemon}

\end{document}